\documentclass[twocolumn,pre,aps,showpacs,floatfix,superscriptaddress]{revtex4}
\usepackage{epsfig}
\usepackage{graphicx}
\begin{document}

\widetext
\title{New approach to the treatment of separatrix chaos \\
and
%\\
%the facilitation of
its application to
the global chaos onset between adjacent
separatrices}

\author{S.M. Soskin}
\altaffiliation{Also at Physics Department, Lancaster University,
%Lancaster LA1 4YB,
UK} \affiliation{Institute of Semiconductor Physics, National
Academy of Sciences of Ukraine, 03028 Kiev, Ukraine}
\affiliation{Abdus Salam ICTP, 34100 Trieste, Italy}
\author{R. Mannella}
\affiliation{Dipartimento di Fisica, Universit\`{a} di Pisa, 56127
Pisa, Italy}
\author{O.M. Yevtushenko} \affiliation{Abdus Salam
ICTP, 34100 Trieste, Italy} \affiliation{Physics Department,
   Ludwig-Maximilians-Universit{\"a}t M{\"u}nchen,
   D-80333 M{\"u}nchen, Germany}

\begin{abstract}
We have developed the {\it general method} for the description of
{\it separatrix chaos}, basing on the analysis of the separatrix map
dynamics. Matching it with the resonant Hamiltonian analysis, we
show that, for a given amplitude of perturbation, the maximum width
of the chaotic layer in energy may be much larger than it was
assumed before. We apply the above theory to explain the drastic
facilitation of global chaos onset in time-periodically perturbed
Hamiltonian systems possessing two or more separatrices, previously
discovered (PRL 90, 174101 (2003)). The theory well agrees with
simulations. We also discuss generalizations and applications.
Examples of applications of the facilitation include: the increase
of the DC conductivity in spatially periodic structures, the
reduction of activation barriers for noise-induced transitions and
the related acceleration of spatial diffusion, the facilitation of
the stochastic web formation in a  wave-driven or kicked oscillator.
\end{abstract}

\pacs{05.45.-a, 05.45.Ac, 05.45.Pq} \maketitle

\section{INTRODUCTION}

A weak perturbation of a Hamiltonian system causes the onset of
chaotic layers around separatrices of the unperturbed system and/or
separatrices surrounding nonlinear resonances generated by the
perturbation
\cite{Chirikov:79,lichtenberg_lieberman,Zaslavsky:1991,zaslavsky:1998,zaslavsky:2005}.
The system may be transported along the layer in a random-like
fashion and this chaotic transport plays an important role in many
physical phenomena
\cite{Zaslavsky:1991,zaslavsky:1998,zaslavsky:2005}. If the
perturbation is sufficiently weak, then the layers are thin and the
chaos is called {\it local}
\cite{Chirikov:79,lichtenberg_lieberman,Zaslavsky:1991,zaslavsky:1998}.
As the perturbation magnitude increases, the width of the layer
grows and the layers corresponding to adjacent separatrices
reconnect at some, typically non-small, critical value of the
perturbation. This conventionally marks the onset of {\it global}
chaos
\cite{Chirikov:79,lichtenberg_lieberman,Zaslavsky:1991,zaslavsky:1998}
i.e. chaos in a large region of the phase space, with chaotic
transport throughout the whole relevant energy range.

The reconnection of the layers around separatrices of the
resonances often correlates with the overlap in energy between
neighbouring resonances calculated independently in the resonant
approximation. The latter constitutes the heuristic Chirikov
resonance-overlap criterion
\cite{Chirikov:79,lichtenberg_lieberman,Zaslavsky:1991,zaslavsky:1998}.
But the Chirikov criterion may fail if the system is of the {\it
zero-dispersion} (ZD) type \cite{pr} i.e. if the frequency of
eigenoscillations possesses a local maximum or minimum as a
function of its energy (cf. also studies of related maps
\cite{Howard:84,Howard:95} which are called {\it nontwist}, {\it
twistless} or {\it nonmonotonic twist} maps). In such systems,
there are typically two resonances of one and the same order
\cite{more}, and their overlap in energy does not result in the
onset of global chaos \cite{pr,Howard:84,Howard:95}. Even their
overlap in phase space \cite{reconnection} results typically only
in the reconnection of the thin chaotic layers associated with the
resonances. As the amplitude of the time-periodic perturbation
grows further, the layers may {\it separate} again
\cite{pr,Howard:84,Howard:95}. An example of the evolution of
resonances in the plane of energy and slow angle is given in
%despite the {\it growth} of the width of the overall relevant range of energy
Fig. 1 (the typical evolution of a real Poincar\'{e} section is
shown e.g. in \cite{comment}).

\begin{figure}[tb]
\includegraphics[width = 5 cm]{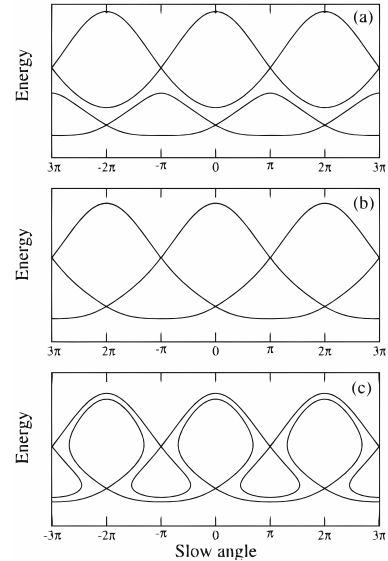}
\caption {Typical evolution of thin chaotic layers in the plane of
slow variables of a zero-dispersion system (the perturbation
magnitude grows from the top to the bottom).}
\end{figure}

As it is known \cite{pr}, any Hamiltonian system with two or more
separatrices belongs to the ZD type: the eigenfrequency as a
function of energy possesses a local maximum between each pair of
adjacent separatrices. For the purpose of global chaos onset, our
letter \cite{prl2003} has addressed the possibility to combine the
overlap of resonances with each other (typical of ZD systems) and
their overlap with the chaotic layers associated with the
separatrices. Via numerical simulations, \cite{prl2003} demonstrated
that this is possible, leading to a scenario for global chaos onset
which requires much smaller perturbation amplitudes than in the
conventional case. The letter \cite{prl2003}
%It also
suggested also a heuristic theory for this effect (more details
were presented in \cite{SPIE}).

The {\it present} work develops the
%first ever
method for the
quantitative description of chaotic layers in {\it phase space}, for
the resonance frequency range. We uncover the physical mechanism of
their overlap with the resonances, and on this basis develop a
detailed {\it self-contained} theory of the facilitated onset of
global chaos. We also discuss generalizations and applications. The
method for the description of the chaotic layers is of  {\it general
importance}: in conventional systems with a single-separatrix layer,
it predicts a much larger maximum width in energy than what was
assumed before
\cite{Chirikov:79,lichtenberg_lieberman,Zaslavsky:1991,zaslavsky:1998,zaslavsky:2005}.

The paper is organized as follows. Sec. II introduces some relevant
model example and presents the major results of the simulations:
studying numerically the frequency dependence of the minimal
amplitude of the AC drive for which global chaos occurs,
$h_{gc}(\omega_f)$, we show that $h_{gc}(\omega_f)$ possesses deep
spikes at certain frequencies. Sec. III gives the self-consistent
asymptotic theory for the minima of the spikes, after assessing the
boundaries of the relevant chaotic layers. Sec. IV gives the theory
for the spikes wings. Discussion of a few generalizations and
applications is carried out in Sec. V. Conclusions are drawn in Sec.
VI. The Appendix describes in details the new method for the
%description
analysis of separatrix chaos.

\section{MODEL AND MAJOR RESULTS OF SIMULATIONS}

As an example of a one-dimensional
Hamiltonian system possessing two or more separatrices, we use a spatially periodic
potential system with two different-height barriers per period (Fig. 2(a)):

\begin{eqnarray}
&&
H_0(p,q)=\frac{p^2}{2}+U(q), \quad\quad U(q)=\frac{(\Phi-\sin(q))^2}{2},
\nonumber\\
&& \Phi={\rm const}<1.
\end{eqnarray}

This model may relate e.g. to a pendulum spinning about its vertical
axis \cite{andronov} or to a classical 2D electron gas in a magnetic
field spatially periodic in one of the in-plane dimensions
\cite{oleg98,oleg99}. The interest to the latter system arose in the
90th due to technological advances allowing to manufacture magnetic
superlattices of high-quality \cite{Oleg12,Oleg10} leading to a
variety of interesting behaviours of the charge carriers in
semiconductors
\cite{oleg98,oleg99,Oleg12,Oleg10,Shmidt:93,shepelyansky}.

Figs. 2(b) and 2(c) show respectively the separatrices of the Hamiltonian system (1) in the $p-q$ plane
and the dependence of the
frequency $\omega$ of its
oscillation, often called {\it eigenfrequency}, on its energy $E\equiv H_0(p,q)$.
%As seen from the latter,
The separatrices correspond to energies equal to the value of the potential
barrier tops $E_b^{(1)}\equiv (1-\Phi)^2/2$ and
$E_b^{(2)}\equiv (1+\Phi)^2/2$ (Fig. 2(a)). The function
$\omega(E)$ is close to the extreme eigenfrequency
$\omega_m\equiv\omega(E_m)$ for most of the range
$[E_b^{(1)},E_b^{(2)}]$ while sharply decreasing to zero as $E$
approaches either $E_b^{(1)}$ or $E_b^{(2)}$.
%Such features of $\omega(E)$ are {\it typical} for
%systems with two or more separatrices.

\begin{figure}[tb]
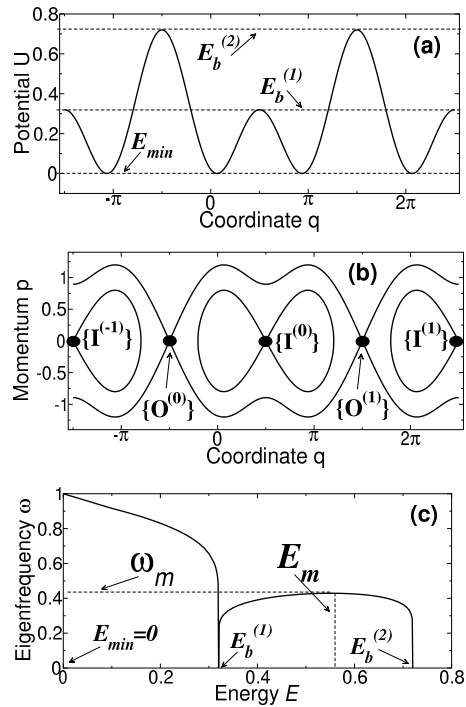

\includegraphics[width = 6 cm]{prefig2a.eps}
\vskip 0.2 cm
\includegraphics[width = 6 cm]{prefig2b.eps}
\vskip 0.2 cm
\includegraphics[width = 6 cm]{prefig2c.eps}
\vskip 0.2 cm
\caption {The potential $U(q)$, the separatrices in the phase
space and the eigenfrequency $\omega(E)$ for the unperturbed
system (1) with $\Phi=0.2$, in (a), (b) and (c) respectively. }
\end{figure}

Add now a time-periodic perturbation: as an example, we use
an AC drive, which corresponds to a dipole \cite{Zaslavsky:1991,Landau:76}
perturbation of the Hamiltonian:
\begin{eqnarray}
&&
\dot{q} = \partial H/\partial p, \quad\quad \dot{p} = -\partial
H/\partial q,
%\dot{q} = \partial H/\partial p, \quad\quad \dot{p} = -\partial H/\partial q,
\\
&&
H(p,q)=H_0(p,q)- h q\cos (\omega_f t).
\nonumber
\end{eqnarray}

The {\it conventional} scenario of global chaos onset
between the separatrices of
the system (2)-(1) is illustrated by Fig.\ 3. The figure presents
the evolution of the stroboscopic Poincar\'{e} section as $h$
grows while $\omega_f$ is fixed at an arbitrarily chosen value
{\it away} from $\omega_m$ and its harmonics. At
small $h$, there are two thin chaotic layers around the inner and
outer separatrices of the unperturbed system.
%The layers are {\it homogeneous} on a coarse-grain scale.
Unbounded chaotic transport takes place only in the outer chaotic
layer i.e.\ in a {\it narrow} energy range. As $h$ grows, so do also
the layers.
%, still remaining homogeneous on the coarse-grain scale.
At some critical value $h_{gc} \equiv h_{gc} (\omega_f)$, the layers
merge. This may be considered as the onset of global chaos: the
whole range of energies between the barrier levels is involved, with
unbounded chaotic transport. The states $\{I^{(l)} \}\equiv
\{p=0,q=\pi/2 +2\pi l \}$ and $\{O^{(l)} \}\equiv\{p=0,q=-\pi/2
+2\pi l\}$ (where $l$ is any integer) in the
%(for $t=n2\pi/\omega_f$ with $n=0,1,2,...$)
Poincar\'{e} section are associated respectively with the inner
and outer saddles of the unperturbed system, and necessarily
belong to the inner and outer chaotic layers, respectively. Thus,
the necessary and sufficient condition for global chaos onset
may be formulated as the possibility for the system placed
initially in the state $\{I^{(0)} \} $ to pass beyond the
neighbouring of the \lq\lq outer'' states, $\{O^{(0)} \} $ or $\{O^{(1)}
\} $, i.e. the coordinate $q$ becomes $<-\pi/2$ or
$>3\pi/2$ at sufficiently large times $t\gg 2\pi/\omega_f$.

\begin{figure}[tb]
\includegraphics[width = 6 cm]{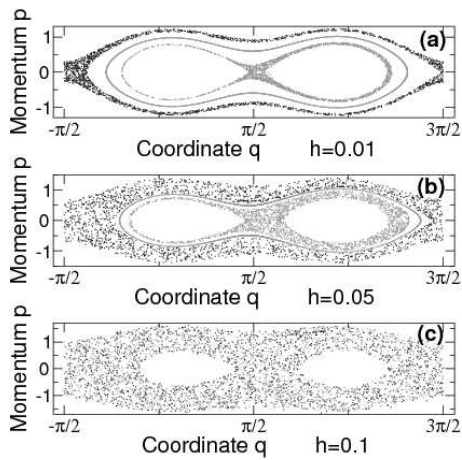}
\caption{The evolution of the stroboscopic (at $t=n2\pi/\omega_f$
with $n=0,1,2,...$) Poincar\'{e} section of the system (2)-(1) with
$\Phi=0.2$ as $h$ grows while $\omega_f=0.3$. The number of points
in each trajectory is 2000. In (a) and (b), three characteristic
trajectories are shown: the inner trajectory starts from the state
$\{I^{(0)} \} \equiv \{p=0,q=\pi/2 \}$ and is chaotic but bounded in
space; the outer trajectory starts from $\{O^{(0)} \}
\equiv\{p=0,q=-\pi/2 \}$ and is chaotic and unbounded in coordinate;
the third trajectory is an example of a regular trajectory
separating the two chaotic ones. In (c), the chaotic trajectories
mix.}
\end{figure}

A diagram in the $h-\omega_f$ plane,  based on the above
criterion, is shown in Fig.\ 4. The lower boundary of the shaded
area represents the function $h_{gc} (\omega_f)$. It has deep {\it spikes}
i.e. cusp-like local minima. The most pronounced spikes are
situated at frequencies $\omega_f=\omega_s^{(j)}$ that are
slightly less than the odd multiples of $\omega_m$,

\begin{equation}
\omega_s^{(j)} \approx\omega_m(2j-1), \quad\quad j=1,2,...
\end{equation}

\noindent The deepest minimum occurs at
$\omega_s^{(1)}\approx\omega_m$: the value of $h_{gc}$ in the
minimum, $h_s^{(1)}\equiv h_{gc} (\omega_ s^{(1)})$, is
approximately 40 times smaller than the value in the neighbouring
pronounced local maximum of $h_{gc} (\omega_f)$ at
$\omega_f\approx 1$. As $n$ increases, the $n$th minimum becomes
less deep. The function $h_{gc} (\omega_f)$ is very sensitive to
$\omega_f$ in the vicinity of the minima: for example, a shift of
$\omega_f$ down from $\omega_s^{(1)}\approx 0.4$ by only 1\%
causes an increase of $h_{gc}$ by $\approx 30\%$.

\begin{figure}[tb]
\includegraphics[width = 7. cm]{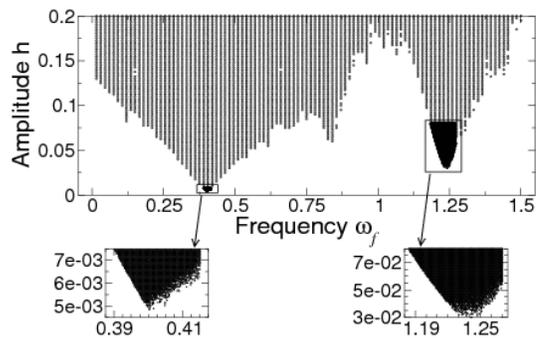}
\caption{The diagram indicating (shade) the perturbation
parameters range for which global chaos exists. Integration time
for each point of the grid is $12000\pi$. }
\end{figure}

The origin of the spikes becomes more clear looking at the
evolution of the Poincar\'{e} section for $\omega_f \approx
\omega_s^{(1)}$ as $h$ grows (Fig. 5): it drastically differs from the
conventional evolution shown in Fig. 3. For $h=0.001$ (Fig.\ 5(a)), one can
see four chaotic trajectories. Two of them are associated with the inner and
outer separatrices of the unperturbed system, similarly to the conventional case (cf.
Fig. 3).
%\cite{Chirikov:79,lichtenberg_lieberman,Zaslavsky:1991}.
They are
marked by green and blue respectively.
%Below, these trajectories will be referred to as ``the inner and outer chaotic layers''.
These trajectories fill the corresponding chaotic layers, which
will be referred below as the \lq\lq inner" and \lq\lq outer"
separatrix layers respectively. The other two chaotic trajectories
marked by red and cyan are associated with the two nonlinear
resonances of the 1st order.
%\cite{Chirikov:79,lichtenberg_lieberman,Zaslavsky:1991}.
Examples of non-chaotic trajectories separating the chaotic ones
are shown in brown. As the perturbation amplitude $h$ increases,
the outer separatrix layer sequentially absorbs other chaotic
trajectories while large stability islands (associated with the
resonances) arise in the layer. At $h=0.003$, it has absorbed the
red trajectory: the resulting chaotic layer is shown in blue in
Fig. 5(b). At $h=0.00475$, this chaotic layer has absorbed the cyan
chaotic trajectory: the resulting chaotic layer is shown in blue
in Fig. 5(c) \cite{20_prime}. Finally, at $h=0.0055$ the latter
blue layer has merged with the inner separatrix layer
\cite{footnote2} (see Fig.\ 5(d)), i.e. the onset of global chaos
as defined above has occurred.

\begin{figure}[tb]
\includegraphics[width = 7. cm]{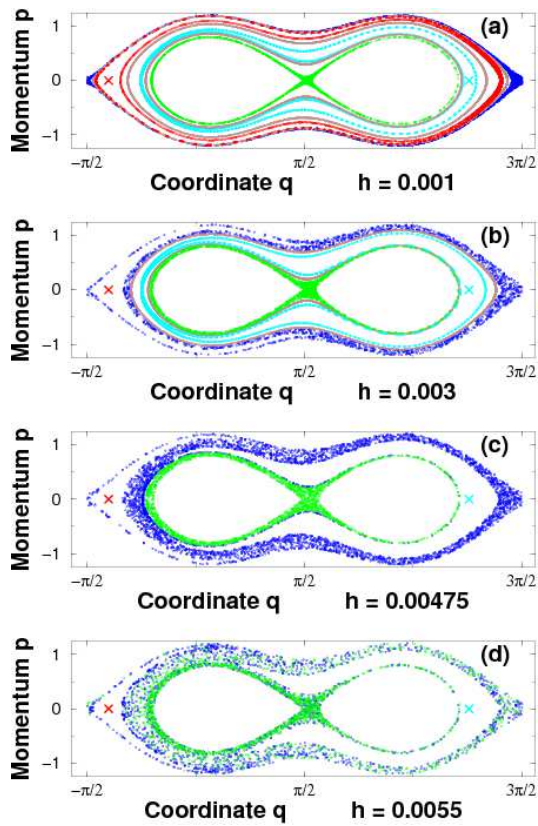}
\caption{The evolution of the stroboscopic Poincar\'{e} section of
the system (2)-(1) with $\Phi=0.2$ as the amplitude of the
perturbation $h$ grows while the frequency is fixed at
$\omega_f=0.401$. The number of points in each trajectory is 2000.
The chaotic trajectories starting from the states $\{I^{(0)} \} $
and $\{O^{(0)} \} $ are drawn in green and blue respectively. The
stable stationary points of Eq. (14) (the 1st-order nonlinear
resonances) are indicated by the red and cyan crosses. The chaotic
layers associated with the resonances are indicated in red and
cyan respectively, unless they merge with those associated
with the green/blue chaotic trajectories. Examples of regular
trajectories embracing the state $\{I^{(0)} \} $ while separating
various chaotic trajectories are shown in brown.}
\end{figure}

\begin{figure}[tb]
\includegraphics[width = 6. cm]{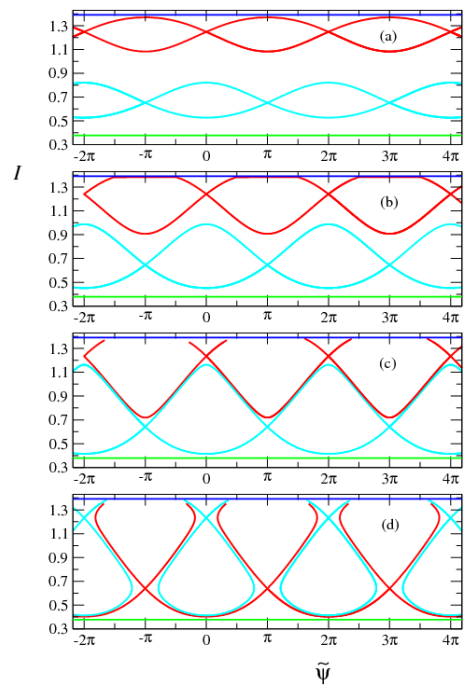}
\caption{The evolution of the separatrices (full lines) of the
1st-order resonances within the resonance approximation (described by
%the Hamiltonian
(4) with $n=1$) in the plane of action $I$ and
slow angle $\tilde{\psi}$, for the same parameters as in Fig.\ 5.
Horizontal levels mark the values of $I$ corresponding to the barriers.}
\end{figure}

Even prior to the theoretical analysis, one can draw a few
conclusions from the evolution. Namely, if
%the frequency
$\omega_f$ is
close to the minimum of the spike of $h_{gc} (\omega_f)$, then

\begin{itemize}

\item[1)] the onset of global chaos occurs due to the {\it
combination} of the overlap of chaotic layers associated with
nonlinear resonances with each other and the overlap of the latter
layers with the inner and outer separatrix layers;

\item[2)] the width of the nonlinear resonances are large
already at quite {\it small} amplitudes of the perturbation, so
that the overlap with the chaotic layers around the original
separatrices occurs at unusually small perturbation amplitudes;

\item[3)] the onset of the overlap of at least one of the
nonlinear resonances with the outer separatrix layer occurs at
values of $h$ which are a few times smaller than those required
for the onset of the overlap with the inner separatrix layer.

\end{itemize}

The above conclusions are also illustrated by Fig. 6 which presents
the evolution of the phase space of slow variables \cite
{Chirikov:79,lichtenberg_lieberman,Zaslavsky:1991,zaslavsky:1998,zaslavsky:2005,pr},
action $I\equiv I(E)$ and slow angle $\tilde{\psi}\equiv
\psi-\omega_ft$, calculated in resonance approximation for the
1st-order spike (see Eq. (4) below).

Similarly, for the spikes of higher order,
higher-order resonances are relevant.

\section{EXPLICIT ASYMPTOTIC THEORY FOR THE MINIMA OF THE SPIKES}

The eigenfrequency $\omega(E)$ is close to its local maximum $\omega_m$
for most of the relevant range $[E_b^{(1)},E_b^{(2)}]$
(Fig. 2(c)). As shown below, $\omega(E)$
approaches a {\it rectangular} form in the asymptotic limit $\Phi
\rightarrow 0$. Hence, if the perturbation frequency $\omega_f $
is close to $\omega_m$ or its odd multiples, $|\omega_f - (2j-1)
\omega_m | \ll \omega_m $, then the energy width of
nonlinear resonances becomes comparable to the width of the
whole range between barriers (i.e. $ E_b^{(2)}-E_b^{(1)}
\approx 2\Phi $)
%already
at a rather small perturbation magnitude $ h \ll \Phi $. Note that
$\Phi $ determines the characteristic magnitude of the perturbation
required for the conventional overlap of the separatrix chaotic
layers, when $\omega_f$ is not close to any odd multiple of
$\omega_m $ (Fig. 3 (c)). Thus, if $ \omega_f\approx \omega_s^{(j)}
$, the nonlinear resonances should play a crucial role in the onset
of global chaos (cf. Fig. 5).

We note that it is not entirely obvious {\it a priori} whether it is
indeed possible to calculate $h_s^{(j)}\equiv
h_{gc}(\omega_s^{(j)})$ within the resonance approximation: in fact,
it is essential for the separatrices of nonlinear resonances to
nearly touch the barriers levels, but the resonance approximation is
obviously invalid in the close vicinity of the barriers;
furthermore, numerical calculations of resonances show that, if
$\omega_f\approx\omega_s^{(j)}$, the perturbation amplitude $h$ at
which the resonance separatrix touches a given energy level in the
close vicinity of the barriers is very sensitive to $\omega_f$,
apparently making the calculation of $h_s^{(j)}$ within the
resonance approximation even less feasible.

Nevertheless, we show below in a self-consistent manner that, in
the asymptotic limit $\Phi\rightarrow 0$, the relevant boundaries of
the chaotic layers lie in the range of energies $E$ where
$\omega(E)\approx\omega_m$. Therefore, the resonant approximation is
valid and it allows to obtain {\it explicit} asymptotic expressions
both for $\omega_s^{(j)} $ and $h_s^{(j)}$, and for the wings of
the spikes in the vicinities of $\omega_s^{(j)}$.

The {\it asymptotic} limit $\Phi\rightarrow 0$ is the most
interesting one from a theoretical point of view since this limit
leads to the strongest facilitation of the global chaos onset and it
is most accurately described by the self-contained theory. Most of
the theory presented below assumes this limit and concentrates
therefore on the results to the {\it lowest} order in the small
parameter.
%, referred below as the {\it leading-order} results.

On the applications side, the range of  {\it moderately small}
$\Phi$  is more interesting, since the chaos facilitation is still
pronounced (and still described by the asymptotic theory) while the
area of chaos between the separatrices is not too small (comparable
with the area inside the inner separatrix): cf. Figs. 2, 3 and 5. To
increase the accuracy of the theoretical description in this range,
we estimate the next-order corrections and develop an efficient
numerical procedure allowing for the further corrections.

\subsection{Resonant Hamiltonian and related quantities}

Let $\omega_f$ be close to the $n$th odd \cite{even} harmonic of
$\omega_m$, $n \equiv (2j-1)$. Over most of the range
$[E_b^{(1)},E_b^{(2)}]$, except in the close vicinities of
$E_b^{(1)}$ and $E_b^{(2)}$, the $n$th harmonic of eigenoscillation
is nearly resonant with the perturbation. Due to this, the (slow)
dynamics of the action $I\equiv I(E) =(2\pi)^{-1}\oint dqp$ and the
angle $\psi$
\cite{Chirikov:79,lichtenberg_lieberman,Zaslavsky:1991,zaslavsky:1998,zaslavsky:2005,pr,Howard:84,Howard:95,Landau:76}
can be shown to be described by the following auxiliary Hamiltonian
(cf.
\cite{Chirikov:79,lichtenberg_lieberman,Zaslavsky:1991,zaslavsky:1998,zaslavsky:2005,pr,Howard:84,Howard:95}):

\begin{eqnarray}
&& \tilde{H}(I,\tilde{\psi})=\int_{I(E_m)}^{I}{\rm d}\tilde{I}\;
(n\omega-\omega_f)\;-\; nhq_n\cos(\tilde{\psi})
\\
&& \quad\; \equiv\; n(E-E_m)-\omega_f(I-I(E_m))\;-\;
nhq_n\cos(\tilde{\psi})\;, \nonumber
\\
&& I \equiv I(E) = \int_{E_{\rm min}}^E
\frac{{\rm d}\tilde{E}}{\omega(\tilde{E})}, \quad\quad  E \equiv H_0(p,q),
\nonumber
\\
&& \tilde{\psi}=n\psi-\omega_ft, \quad\quad \nonumber
\\
&& \psi= \pi+{\rm sgn}(p)\omega(E)\int^q_{q_{\rm
min}(E)}\frac{{\rm d}\tilde{q}}{\sqrt{2(E-U(\tilde{q}))}}+2\pi l,
\nonumber
\\
&& q_n\equiv q_n(E)= \frac{2}{\pi}\int_0^{\pi/2} \!\!\!\!\! {\rm
d}\psi \; q(E,\psi)\cos(n\psi) ,
%\quad i\equiv \sqrt{-1},
\nonumber
\\
&& |n\omega-\omega_f|\ll\omega,\quad\quad n\equiv 2j-1, \quad\quad
j=1,2,3,\ldots, \nonumber
\end{eqnarray}

\noindent where $E_{\rm min}$ is the minimal energy
 (over all $q,p$)
$ E \equiv H_0(p,q)$; $\omega\equiv \omega(E) =d H_0/d I$ and
$q_{\rm min}(E)$  are, respectively, the frequency and the minimal
coordinate of the conservative motion with a given value of energy
$E$; $l$ is the number of right turning points in the trajectory
$[q(\tau)]$ of the conservative motion with energy $E$ and given
initial state $(q_0,p_0)$.

Let us derive the explicit expressions for various quantities in
(4). In the unperturbed case ($h=0$), the equations of motion (2)
with $H_0$ (1) can be integrated \cite{oleg99} (see also Eq. (60)
below), so that we can find $\omega(E)$:

\begin{eqnarray}
&&
\omega(E)=\frac{\pi(2E)^{1/4}}{2K
\left[k \right]},
\\
&&
k=\frac{1}{2}\sqrt{\frac{(\sqrt{2E}+1)^2-\Phi^2}{\sqrt{2E}}} \, ,
\nonumber
\end{eqnarray}

\noindent
where

\begin{equation}
K[k]=\int_0^{\frac{\pi}{2}}\frac{{\rm d}\phi}{\sqrt{1-k^2\sin^2(\phi)}},
\end{equation}

\noindent
is the full elliptic integral of the first order \cite{Abramovitz_Stegun}. Using
its asymptotic expression,
%for $ K[k] $ \cite{Abramovitz_Stegun},

\[
K[k\rightarrow 1] \simeq \frac{1}{2} \ln\left( { 16 \over 1-k^2 }
\right),
\]

\noindent we derive $\omega(E)$ in the asymptotic limit $\Phi
\rightarrow 0$:

\begin{eqnarray}
&& \omega(E) \simeq \frac{\pi}{\ln
\left(
\frac{64}{(\Phi-\Delta E)(\Phi+\Delta E)}
\right)},
\\
&&
\Delta E\equiv E-\frac{1}{2},\quad\quad |\Delta E|<\Phi,
\nonumber
\\
&&
\Phi \rightarrow 0.
\nonumber
\end{eqnarray}

The function $\omega(E)$ (7) is close to its maximum

\begin{equation}
\omega_m\equiv \max_{[E_b^{(1)},E_b^{(2)}]}\{\omega(E)\}
       \simeq \frac{\pi}{2\ln (8/\Phi)} \,
\end{equation}

\noindent for most of the interbarrier \cite{barriers} range of
energies $[1/2-\Phi,1/2+\Phi]$; on the other hand, in the close
vicinity of the barriers, where either $|\ln(1/(1-\Delta E/\Phi))|$
or $|\ln(1/(1+\Delta E/\Phi))|$ become comparable with, or larger
than, $\ln(8/\Phi)$, $\omega(E)$ sharply decreases to zero as
$|\Delta E|\rightarrow \Phi$. The range where this takes place is
$\sim\Phi^2$, and its ratio to the whole interbarrier range,
$2\Phi$, is $\sim\Phi$ i.e. it goes to zero in the asymptotic limit
$\Phi\rightarrow 0$: in other words, $\omega(E)$ approaches a {\it
rectangular} form. As it will be clear from the following, {\it\bf
it is this almost rectangular form of $\omega(E)$ which determines
many of the characteristic features of the global chaos onset in
systems with two or more separatrices}.

One more quantity which strongly affects $(\omega_s, h_s)$ is the
Fourier harmonic $q_n\equiv q_n(E)$. The system stays most of the
time very close to one of the barriers. Consider the motion within
one of the periods of the potential $U(q)$, between neighboring
upper barriers $[q_{ub}^{(1)},q_{ub}^{(2)}]$ where
$q_{ub}^{(2)}\equiv q_{ub}^{(1)}+2\pi$. If the energy $E\equiv
1/2+\Delta E$ lies in the relevant range $[E_b^{(1)},E_b^{(2)}]$,
then the system will stay close to the lower barrier $q_{lb}\equiv
q_{ub}^{(1)}+\pi$ for a time \cite {arbitrary_constant}

\begin{equation}
T_l\approx 2\ln \left( \frac{1}{\Phi+\Delta E} \right)
\end{equation}

\noindent
during each period of eigenoscillation, while it will stay close to one of the upper
barriers $q_{ub}^{(1,2)}\equiv q_{lb}\pm \pi$
for most of the remaining of the eigenoscillation,

\begin{equation}
T_u\approx 2\ln \left( \frac{1}{\Phi-\Delta E} \right)\quad .
\end{equation}

\noindent Hence, the function $q(E,\psi)-q_{lb}$ may be
approximated by the following piecewise even periodic function:

\begin{eqnarray}
&& q(E,\psi)-q_{lb}= \left\{_{0\quad {\rm at} \quad \psi\in
\left.\right]\frac{\pi}{2}\frac{T_u}{T_l+T_u},\pi-\frac{\pi}{2}\frac{T_u}{T_l+T_u}
\left[\right.,}^{\pi\quad {\rm at} \quad \psi\in
\left[0,\frac{\pi}{2}\frac{T_u}{T_l+T_u} \right]\cup
\left[\pi-\frac{\pi}{2}\frac{T_u}{T_l+T_u},\pi \right], } \right.
\nonumber
\\
&& q(E,-\psi)-q_{lb}=q(E,\psi)-q_{lb}, \nonumber
\\
&& q(E,\psi\pm 2\pi i)=q(E,\psi),
\nonumber\\
&& i=1,2,3,... \nonumber
\end{eqnarray}

\noindent
Substituting the above approximation for $q(E,\psi)$ into the
definition of $q_n$ (4), one can obtain:

\begin{eqnarray}
&& q_{2j-1}\equiv q_{2j-1}(E)=\frac{2}{2j-1}\sin \left(
\frac{(2j-1)\pi/2}{1+\frac{\ln \left( \frac{1}{\Phi+\Delta E}
\right) }{\ln \left( \frac{1}{\Phi-\Delta E} \right)}} \right) \;,
\nonumber\\
&& \Phi\rightarrow 0, \nonumber
\\
&&
q_{2j}=0,\\
&&
j=1,2,3,...
\nonumber
\end{eqnarray}

At barrier energies, $q_{2j-1}$ takes the values
\[
  q_{2j-1}(E_b^{(1)})=0, \quad\quad q_{2j-1}(E_b^{(2)}) = -(-1)^{j}{ 2 \over (2j-1) } \, .
\]

As $E$ varies in between the barrier values, $q_{2j-1}$ varies
monotonously if $j=1$ and non-monotonously otherwise (cf. Fig. 11).
But in any case, the significant variations occur mostly in the
close vicinity of the barrier energies $E_b^{(1)}$ and $E_b^{(2)}$
while, for most of the range $[E_b^{(1)},E_b^{(2)}]$, the argument
of the sine in Eq. (11) is close to $\, \pi/4$ and $\, q_{2j-1}$ is
then almost constant:

\begin{eqnarray}
&& q_{2j-1}\approx
(-1)^{\left[\frac{2j-1}{4}\right]}\frac{\sqrt{2}}{2j-1}, \quad \
j= 1, 2, 3, \, \ldots,
\\
&& \left| \ln \left( \frac{1+\Delta E/\Phi}{1-\Delta E/\Phi}
\right) \right| \ll 2\ln \left( \frac{1}{\Phi} \right), \nonumber
\end{eqnarray}

\noindent where $[\ldots]$ means the integer part.

In the asymptotic limit $\Phi\rightarrow 0$, the range of $\Delta
E$ where the approximate equality (12) for $q_{2j-1}$ is valid
approaches the whole range $]-\Phi,\Phi[$.

We emphasize that $|q_n|$ determines the \lq\lq strength'' of the
nonlinear resonances: therefore, apart from the nearly
rectangular form of $\omega(E)$, {\bf the non-smallness of $|q_n|$
is one more factor giving rise to the strong facilitation of
the global chaos onset}.

We shall need also the
asymptotic expression of the action $I$. Substituting
$\omega(E)$ (7) into the definition of $I(E)$ (4) and
carrying out the integration, we obtain

\begin{eqnarray}
&& I(E)=I(1/2)+ \frac {\Delta E \ln \left( \frac{64{\rm
e}^2}{\Phi^2-(\Delta E)^2} \right)  + \Phi \ln \left(
\frac{\Phi-\Delta E}{\Phi+\Delta E} \right)}{\pi}
  \; , \nonumber
\\
&&
\Phi\rightarrow 0.
\end{eqnarray}

\subsection{Reconnection of resonance separatrices}

We now turn to the analysis of the {\it phase space} of the
resonance Hamiltonian (4). The evolution of the Poincar\'{e} section
(see Fig. 5 and the related analysis in Sec. II) suggests that we
need to find such {\it separatrix} of (4) which undergoes the
following evolution as $h$ grows: for sufficiently small $h$, the
separatrix does not overlap chaotic layers associated with the
barriers while, for $h>h_{gc}(\omega_f)$, it does overlap them. The
relevance of such a condition will be justified further.

For $\omega_f\approx n\omega_m$ with a given odd $n$, the equations
of motion of the system (4) read as follows

\begin{eqnarray}
&&
\dot{I}=-\frac{\partial \tilde{H}}{\partial \tilde{\psi}}\equiv
-nhq_n\sin(\tilde{\psi}),
\\
&&
\dot{\tilde{\psi}}=\frac{\partial \tilde{H}}{\partial I}\equiv
n\omega - \omega_f-nh\frac{{\rm d}q_n}{{\rm
d}I}\cos(\tilde{\psi}).
\nonumber
\end{eqnarray}

\noindent Any separatrix necessarily includes one or more unstable
stationary points. The system (14) may have several stationary
points per $2\pi$ interval of $\tilde{\psi}$. Let us first exclude
those points which are irrelevant to the separatrix undergoing the
evolution described above.

Given that $q_n(E_b^{(1)})=0$, there are two unstable stationary
points with $I$ corresponding to $E=E_b^{(1)}$ and
$\tilde{\psi}=\pm\pi/2$. They are irrelevant since, even for an
infinitely small $h$, each of them necessarily lies inside the
corresponding barrier chaotic layer.

If $E\neq E_b^{(1)}$, then $q_n\neq 0$, so $\dot{I}=0$ only if
$\tilde{\psi}$ is equal either to 0 or to $\pi$. Substituting these
values into the second equation of (14) and putting
$\dot{\tilde{\psi}}=0$, we obtain the equations for the
corresponding actions:

\begin{equation}
 X_{\mp}(I) \equiv  n\omega-\omega_f\mp nh{\rm d}q_n/{\rm d}I =0,
\end{equation}

\noindent where the signs \lq\lq -" and \lq\lq +" correspond to
$\tilde{\psi}=0$ and $\tilde{\psi}=\pi$ respectively. A typical
example of the graphic solution of equations (15) for $n=1$ is shown
in Fig. 7. Two of the roots corresponding to $\tilde{\psi}=\pi$ are
very close to the barrier values of $I$ (we remind that the relevant
values of $h$ are small). These roots arise due to the divergence of
${\rm d} q/{\rm d} I$ as $I$ approaches any of the barrier values.
The lower/upper root corresponds to a stable/unstable point.
However, for any $n$, both these points and the separatrix generated
by the unstable point necessarily lie in the ranges covered by the
barrier chaotic layers. Therefore, they are also irrelevant
\cite{25_prime}. For $n>1$, the number of the roots of (15) in the
vicinity of the barriers may be larger (due to the oscillations of
the modulus and sign of ${\rm d}q_n/{\rm d}I$ in the vicinity of the
barriers) but they all are irrelevant for the same reason, at least
to leading-order terms in the expressions for the spikes minima.

\begin{figure}[tb]
\includegraphics[width = 6. cm]{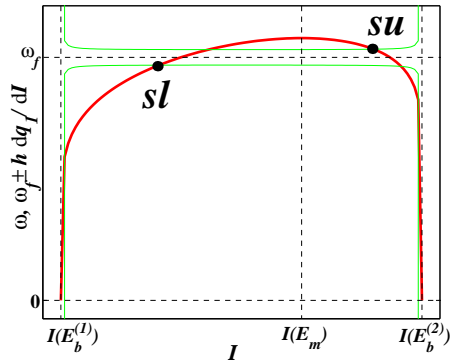}
\caption{A schematic example illustrating the graphic solutions of
Eqs. (15) for $n=1$, as intersections of the curve $\omega(I)$
(thick solid red line) with the curves $\omega_f\pm h{\rm
d}q_n(I)/{\rm d}I$ (thin solid green lines). The solutions
corresponding to the lower and upper relevant saddles (defined by
Eq.(16)) are marked by dots and by the labels $sl$ and $su$
respectively (we do not mark other solutions because they are
irrelevant).}
\end{figure}

Consider the stationary points corresponding to the remaining four
roots of equations (15). As follows from the analysis of equations
(14) linearized near the stationary points (cf.
\cite{Chirikov:79,lichtenberg_lieberman,Zaslavsky:1991,zaslavsky:1998,zaslavsky:2005,pr}),
two of them are stable elliptic points \cite{crosses}, often called
{\it nonlinear resonances}, while two others are unstable hyperbolic
points, often called {\it saddles}. These saddles are of main
interest in the context of our work. They belong to the {\it
separatrices} separating the regions of oscillations around the
resonances from the regions of motion with a \lq\lq running'' slow
angle $\tilde{\psi}$.

We shall distinguish the relevant saddles as the saddles with the
{\it lower} action/energy (using the subscript \lq\lq$sl$'') and
{\it upper} action/energy (using the subscript \lq\lq$su$''). The
positions of the saddles in the $I-\tilde{\psi}$ plane are defined
by the following equations (cf. Figs. 6 and 7):

\begin{eqnarray}
&& g\equiv {\rm
sign}(q_n(I_{su,sl}))=(-1)^{\left[\frac{n}{4}\right]} ,
\\
&& \tilde{\psi}_{sl}=\pi(1+g)/2, \quad\quad
\tilde{\psi}_{su}=\pi(1-g)/2,\nonumber
\\
&& X_g(I_{sl})=X_{-g}(I_{su})=0, \nonumber
\\
&& \frac {{\rm d}X_g(I_{sl})}{{\rm d}I_{sl}}>0, \quad\quad \frac
{{\rm d}X_{-g}(I_{su})}{{\rm d}I_{su}}<0, \nonumber
\end{eqnarray}

\noindent where $X_{\pm}(I)$ are defined in Eq. (15) while
$I_{sl}$ and $I_{su}$ are closer to $I(E_m)$ than any other
solution of (16) (if any) from below and from above, respectively.

Given that the values of $h$ relevant to the minima of the spikes are
small in the asymptotic limit $\Phi\rightarrow 0$, one may neglect
the last term in the definition
of $X_{\mp}$ in Eq. (15) in the lowest-order approximation,
so that the equations $X_{\mp}=0$ reduce
to the simple resonance condition

\begin{equation}
n\omega(I_{su,sl})= \omega_f.
\end{equation}

\noindent Substituting here Eq. (7) for $\omega$, we obtain the
explicit expressions for the energies in the saddles:

\begin{eqnarray}
&&
E_{su,sl}\approx \frac{1}{2}\pm \Delta E^{(1)},
\\
&& \Delta E^{(1)}\equiv
\sqrt{\Phi^2-64\exp\left(-\frac{n\pi}{\omega_f}\right)}, \quad\quad
\omega_f\leq n\omega_m. \nonumber
\end{eqnarray}

\noindent
The corresponding actions
$I_{su,sl}$ are expressed via $E_{su,sl}$ by means of Eq. (13).

For $\omega_f \approx n \omega_m $, the values of
$E_{su,sl}$ (18) lie in the range where the expression (12) for
$q_n$ does hold true. This will be explicitly confirmed by the
results of the calculations based on this assumption.

Using (16) for the angles and (18) for the energies, and the
asymptotic expressions (7), (12) and (13) for $\omega(E)$, $q_n(E)$
and $I(E)$ respectively, and allowing for the resonance condition
(17), we obtain explicit expressions for the values of the
Hamiltonian (4) in the saddles:

\begin{equation}
\tilde{H}_{sl}=-\tilde{H}_{su}=\frac{\omega_f}{\pi}
\left[
2\Delta E^{(1)}-\Phi\ln
\left(\frac{\Phi+\Delta E^{(1)}}{\Phi-\Delta E^{(1)}}
\right)
\right]
+h\sqrt{2}.
\end{equation}

As the analysis of simulations suggests (see the item 1 in the end
of Sec. II) and as it is rigorously shown in the next subsection,
one of the main conditions which should be satisfied in the spikes
is the overlap in phase space between the separatrices of the
nonlinear resonances, called {\it separatrix reconnection}
\cite{pr,Howard:84,Howard:95}. Given that the Hamiltonian
$\tilde{H}$ is constant along any trajectory of the system (4), the
values of $\tilde{H}$ in the lower and upper saddles of the {\it
reconnected} separatrices are equal to each other:

\begin{equation}
\tilde{H}_{sl}=\tilde{H}_{su} \,,
\end{equation}

\noindent that may be considered as the necessary and sufficient
\cite{27} condition for the reconnection. Taking into account that
$\tilde{H}_{sl}=-\tilde{H}_{su}$ (see (19)), it follows from (20)
that

\begin{equation}
\tilde{H}_{sl}=\tilde{H}_{su}=0.
\end{equation}

Explicitly, the relations in (21) reduce to
%the following one:

\begin{eqnarray}
&& h\equiv h(\omega_f)=\frac{\omega_f}{\sqrt{2}\pi} \left[\Phi\ln
\left(\frac{\Phi+\Delta E^{(1)}}{\Phi-\Delta E^{(1)}} \right)
-2\Delta E^{(1)} \right], \nonumber
\\
&& \Delta E^{(1)}\equiv
\sqrt{\Phi^2-64\exp(-\frac{n\pi}{\omega_f})},
\\
&&
0<\omega_m-\omega_f/n\ll \omega_m\equiv \frac{\pi}{2\ln(8/\Phi)},
\nonumber
\\
&&
n=1,3,5,...
\nonumber
\end{eqnarray}

\noindent The function $h(\omega_f)$ (22) monotonously decreases to
zero as $\omega_f$ grows from $0$ to $n\omega_m$, where the line
abruptly stops. Fig. 10 shows the portions of the lines (22)
relevant to the left wings of the 1st and 2nd spikes (for
$\Phi=0.2$).

\subsection {Barrier chaotic layers}

The next step is to find a minimal value of $h$ for which the
resonance separatrix overlaps the chaotic layer related to a
potential barrier. With this aim, we study how the relevant outer
boundary of the chaotic layer behaves as $h$ and $\omega_f$ vary.
Assume that the relevant $\omega_f$ is close to $n\omega_m$ while
the relevant $h$ is sufficiently large for $\omega(E)$ to be close
to $\omega_m$ at all points of the outer boundary of the layer (the
results will confirm these assumptions). Then the motion along the
regular trajectory infinitesimally close to the layer boundary may
be described within the resonance approximation (4). Hence the
boundary may also be described as a trajectory of the resonant
Hamiltonian (4). This is explicitly proved in the Appendix, using
the separatrix map analysis that allows for the validity of the
relation $\omega(E)\approx\omega_m$ for all $E$ relevant to the
boundary of the chaotic layer. The main results are presented below.
For the sake of clarity, we present them for each layer separately,
although they are similar in practice.

\subsubsection{Lower layer}

Let $\omega_f$ be close to any of the spikes minima.

One of the key roles in the formation of the upper boundary of the
layer is played by the angle-dependent quantity $\delta_l|\sin(
\tilde{\psi})|$ which we call the {\it generalized separatrix split}
(GSS) for the lower layer, alluding to the conventional {\it
separatrix split} \cite{zaslavsky:1998} for the lower layer
$\delta_l \equiv |\epsilon^{(low)}( \omega_f)|h$ with
$\epsilon^{(low)}$ given by Eq. (A11) \cite{melnikov}. Accordingly,
we use the term \lq\lq lower GSS curve'' for the following curve in
the $I-\tilde{\psi}$ plane:

\begin{equation}
I=I_{\rm GSS}^{(l)}(\tilde{\psi})\equiv
I(E_b^{(1)}+\delta_l|\sin(\tilde{\psi})|).
\end{equation}

\vskip 0.2cm

\hskip 2.5cm {\it a. Relatively small $h$}

\vskip 0.2cm

If $h < h_{cr}^{(l)} (\omega_f)$, where the critical value $ \,
h_{cr}^{(l)}(\omega_f) \, $ is determined by Eq. (41) (its origin
will be explained further), then there are differences in the
boundary formation for the frequency ranges of {\it odd} and {\it
even} spikes. We describe these ranges separately.

\vskip 0.2cm

\hskip 2.9cm {\it a.1. Odd spikes}

\vskip 0.2cm

In this case, the boundary is formed by the trajectory of the
Hamiltonian (4) {\it tangent} to the GSS curve (see Fig. 16(a); cf.
also Figs. 8(a), 9(b), 9(c)). There are two tangencies in the angle
range $]-\pi,\pi[$: they occur at the angles
$\pm\tilde{\psi}_t^{(l)}$ where $\tilde{\psi}_t^{(l)}$ is determined
by Eq. (A21).

In the ranges of $h$ and $\omega_f$ relevant to the spike minimum,
the asymptotic expressions for $\delta_l$ and
$\tilde{\psi}_t^{(l)}$ are:

\begin{eqnarray}
&&
\delta_l = \sqrt{2}\pi h,
\\
&& \tilde{\psi}_t^{(l)}= (-1)^{\left[\frac{n}{4}\right]}
        \sqrt{\frac{n\pi}{8\ln\left(1/\Phi\right)}} +\pi\frac{1-(-1)^{\left[\frac{n}{4}\right]}}{2}.
\end{eqnarray}
Hence, the asymptotic value for the deviation of the tangency
energy $E_t^{(l)}$ from the lower barrier reduces to:
\begin{equation}
E_t^{(l)}-E_b^{(1)} \equiv \delta_l\sin(\tilde{\psi}_t^{(l)} ) =
\frac{\pi^{3/2}} {2}\frac{h}{\sqrt{\ln\left(1/\Phi\right)/n}}.
\end{equation}

The minimal energy on the boundary, $E_{\min}^{(l)}$, corresponds
to $\tilde{\psi}=$ $0$ or $\pi$ for even or odd values of $[n/4]$
respectively. Thus, it can be found from the equality
\begin{eqnarray}\label{27}
&&
\tilde{H}\left(I(E_{\min}^{(l)}),\tilde{\psi}=\pi(1-(-1)^{\left[\frac{n}{4}\right]})/2\right)=
\nonumber
\\
&& =\tilde{H}\left(I_t^{(l)}\equiv
I(E_t^{(l)}),\tilde{\psi}_t^{(l)}\right).
\end{eqnarray}

At $\Phi\rightarrow 0$, Eq. (\ref{27}) yields the following
expression for the minimal deviation of energy on the boundary from the barrier:
\begin{eqnarray}
\delta_{\min}^{(l)}&\equiv&
E_{\min}^{(l)}-E_b^{(1)}=(E_t^{(l)}-E_b^{(1)}) /\sqrt{{\rm
e}}=\nonumber
\\
&=&\frac{\pi^{3/2}} {2\sqrt{{\rm
e}}}\frac{h}{\sqrt{\ln\left(1/\Phi\right)/n}}.
\end{eqnarray}

In the context of global chaos onset, the most important
property of the boundary is that the {\it maximal} deviation of its
energy from the barrier, $\delta_{\max}^{(l)}$, greatly exceeds both
$\delta_{\min}^{(l)}$ and $\delta_{l}$. As $h\rightarrow
h_{cr}^{(l)}$, the maximum of the boundary approaches the saddle
\lq\lq{\it sl}''.

\vskip 0.2cm

\hskip 2.9cm {\it a.2. Even spikes}

\vskip 0.2cm

In this case, the Hamiltonian (4) possesses saddles \lq\lq {\it s}''
in the close vicinity to the lower barrier (see Fig. 16(b)). Their
angles differ by $\pi$ from those of \lq\lq {\it sl}'':
\begin{eqnarray}
&&\tilde{\psi}_s=\pi\frac{1-(-1)^{\left[\frac{n}{4}\right]}}{2}+2\pi
m,
\\
&& m=0,\pm 1, \pm 2, \ldots, \nonumber
\end{eqnarray}

\noindent while the deviation of their energies from the barrier
still lies in the relevant (resonant) range and reads, in the
lowest-order approximation,

\begin{equation}
\delta_s=\frac{\pi}{2\sqrt{2}}\frac{h}{\ln(\ln(1/\Phi))}.
\end{equation}

The lower whiskers of the separatrix generated by these saddles
intersect the GSS curve while the upper whiskers in the asymptotic
limit do not intersect it (Fig. 16(b)). Thus, it is the upper
whiskers of the separatrix which form  the boundary of the chaotic
layer in the asymptotic limit. The energy on the boundary takes the
minimal  value right on the saddle \lq\lq {\it s}'', so that

\begin{equation}
\delta_{\min}^{(l)}=\delta_s=\frac{\pi}{2\sqrt{2}}\frac{h}{\ln(\ln(1/\Phi))}.
\end{equation}

Similar to the case of the odd spikes, the {\it maximal} (along the
boundary) deviation of the energy from the barrier greatly exceeds
both $\delta_{\min}^{(l)}$ and $\delta_{l}$. As $h\rightarrow
h_{cr}^{(l)}$, the maximum of the boundary approaches the saddle
\lq\lq{\it sl}''.

\vskip 0.2cm

\hskip 2.9cm {\it b. Relatively large $h$}

\vskip 0.2cm

If $h>h_{cr}^{(l)}(\omega_f)$, the previously described trajectory (the
tangent one or the separatrix, for the odd or even spike ranges
respectively) is encompassed by the separatrix of the lower
nonlinear resonance and typically forms the boundary of the major
stability island inside the lower layer (reproduced periodically in
$\tilde{\psi}$ with the period $2\pi$). The upper {\it outer}
boundary of the layer is formed by the upper part of the {\it
resonance separatrix}. This may be interpreted as the absorption of
the lower resonance by the lower chaotic layer.

\begin{figure}[tb]
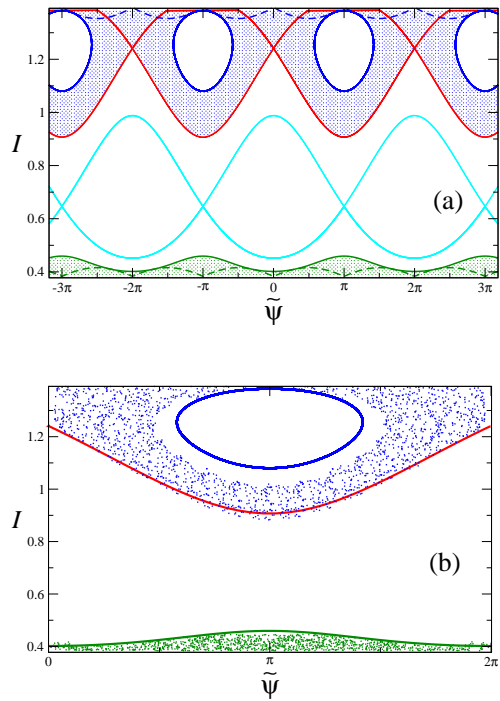

\includegraphics[width = 6.5 cm]{prefig8a.eps}
\vskip 0.75 cm
\includegraphics[width = 6.5 cm]{prefig8b.eps}
\vskip 0.75 cm \caption{(a). Chaotic layers (shaded in green and
blue, for the upper and lower layers respectively) in the plane of
action $I$ and slow angle $\tilde{\psi}$, as described by our
theory. Parameters are the same as in Figs. 5(b) and 6(b). The lower
and upper boundaries of the figure box coincide with $I(E_b^{(1)})$
and $I(E_b^{(2)})$ respectively. The resonance separatrices are
drawn by the cyan and red solid lines (for the lower and upper
resonances respectively). Dashed green and blue lines mark the
curves $I=I_{\rm GSS}^{(l)}(\tilde{\psi})\equiv
I(E=E_b^{(1)}+\delta_l|\sin(\tilde{\psi})|)$ and $I=I_{\rm
GSS}^{(u)}(\tilde{\psi})\equiv
I(E=E_b^{(2)}-\delta_u|\sin(\tilde{\psi})|)$ respectively, where
$\delta_l$ and $\delta_u$ are the values of the separatrix split
related to the lower and upper barrier respectively. The upper
boundary of the lower layer is formed by the trajectory of the
resonant Hamiltonian system (4) tangent to the curve $I=I_{\rm
GSS}^{(l)}(\tilde{\psi})$. The lower boundary of the upper layer is
formed by the lower part of the upper (red) resonance separatrix.
The periodic closed loops (solid blue lines) are the trajectories of
the system (4) tangent to the curve $I_{\rm
GSS}^{(u)}(\tilde{\psi})$: they form the boundaries of the major
stability islands inside the upper chaotic layer. (b). Comparison of
the chaotic layers obtained from computer simulations (dots) with
the theoretically calculated boundaries (solid lines) shown in the
box (a). }
\end{figure}

\subsubsection{Upper layer}

Let $\omega_f$ be close to any of the spikes minima.

One of the key roles in the formation of the lower boundary of the
layer is played by the angle-dependent quantity $\delta_u|\sin(
\tilde{\psi})|$ which we call the {\it generalized separatrix split}
(GSS) for the upper layer; $\delta_u$ is the separatrix split for
the upper layer: $\delta_u = |\epsilon^{(up)}( \omega_f)|h$ with
$\epsilon^{(up)}$ given by Eq. (A43). Accordingly, we use the term
\lq\lq upper GSS curve'' for the following curve in the
$I-\tilde{\psi}$ plane:

\begin{equation}
I=I_{\rm GSS}^{(u)}(\tilde{\psi}) \equiv
I(E_b^{(2)}-\delta_u|\sin(\tilde{\psi})|).
\end{equation}

\vskip 0.2cm

\hskip 2.5cm {\it a. Relatively small $h$}

\vskip 0.2cm

If $h < h_{cr}^{(u)} (\omega_f)$, where the critical value $ \,
h_{cr}^{(u)}(\omega_f) \, $ is determined by Eq. (42) (its origin
will be explained further), then there are some differences in the
boundary formation in the frequency ranges of {\it odd} and {\it
even} spikes: for odd spikes, the formation is similar to the one
for even spikes in the lower-layer case and vice versa.

\vskip 0.2cm

\hskip 2.9cm {\it a.1. Odd spikes}

\vskip 0.2cm

In this case, the Hamiltonian (4) possesses saddles \lq\lq {\it
$\tilde{s}$}'' in the close vicinity to the upper barrier,
analogous to the saddles \lq\lq {\it s}'' near the lower barrier
in the case of even spikes. Their angles are shifted by $\pi$ from
those of \lq\lq {\it s}'':
\begin{eqnarray}
&&\tilde{\psi}_{\tilde{s}}=\tilde{\psi}_s+\pi=\pi\frac{1+(-1)^{\left[\frac{n}{4}\right]}}{2}+2\pi
m,
\\
&& m=0,\pm 1, \pm 2, \ldots \nonumber
\end{eqnarray}

\noindent The deviation of their energies from the upper barrier
coincides, in the lowest-order approximation, with $\delta_{s}$:

\begin{equation}
\delta_{\tilde{s}}
=\delta_s=\frac{\pi}{2\sqrt{2}}\frac{h}{\ln(\ln(1/\Phi))}.
\end{equation}

The upper whiskers of the separatrix generated by these saddles
intersect the upper GSS curve while the lower whiskers in the
asymptotic limit do not intersect it. Thus, it is the lower whiskers
of the separatrix which form  the boundary of the chaotic layer in
the asymptotic limit. The deviation of energy from the upper barrier
takes its minimal (along the boundary) value right on the saddle
\lq\lq {\it $\tilde{s}$}'',

\begin{equation}
\delta_{\min}^{(u)}=\delta_{\tilde{s}}=\frac{\pi}{2\sqrt{2}}\frac{h}{\ln(\ln(1/\Phi))}.
\end{equation}

The {\it maximal} (along the boundary) deviation of the energy from
the barrier greatly exceeds both $\delta_{\min}^{(u)}$ and
$\delta_{u}$. As $h\rightarrow h_{cr}^{(u)}$, the maximum of the
boundary approaches the saddle \lq\lq{\it su}''.

\vskip 0.2cm

\hskip 2.9cm {\it a.2. Even spikes}

\vskip 0.2cm

The boundary is formed by the trajectory of the Hamiltonian (4) {\it
tangent} to the GSS curve. There are two tangencies in the angle
range $]-\pi,\pi[$: they occur at the angles
$\pm\tilde{\psi}_t^{(u)}$ where $\tilde{\psi}_t^{(u)}$ is determined
by Eq. (A41).

In the ranges of $h$ and $\omega_f$ relevant to the spike minimum,
the expressions for $\delta_u$ and $\tilde{\psi}_t^{(u)}$ in the
asymptotic limit $\Phi\rightarrow 0$ are similar to the analogous
quantities in the lower-layer case:
\begin{eqnarray}
&&
\delta_u = \sqrt{2}\pi h,
\\
&& \tilde{\psi}_t^{(u)}=-(-1)^{\left[\frac{n}{4}\right]}
        \sqrt{\frac{n\pi}{8\ln\left(\frac{1}{\Phi}\right)}}
        +\pi\frac{1+(-1)^{\left[\frac{n}{4}\right]}}{2} .
\end{eqnarray}
Hence, the asymptotic value for the deviation of the tangency
energy $E_t^{(u)}$ from the upper barrier reduces to:
\begin{eqnarray}
E_b^{(2)}-E_t^{(u)}&=&\delta_u\left|\pi\frac{1+(-1)^{\left[\frac{n}{4}\right]}}{2}-\tilde{\psi}_t^{(u)}\right|=
\nonumber
\\
&=&\frac{\pi^{3/2}} {2}\frac{h}{\sqrt{\ln\left(1/\Phi\right)/n}}.
\end{eqnarray}

The maximal energy on the boundary, $E_{\max}^{(u)}$, corresponds
to $\tilde{\psi}=\pi(1+(-1)^{[n/4]})/2$. Thus, it can be found
from the equality
\begin{eqnarray}
&&\tilde{H}(I=I(E_{\max}^{(u)}),\tilde{\psi}=\pi(1+(-1)^{[n/4]})/2)=
\nonumber \\
&& = \tilde{H}(I_t^{(u)}\equiv I(E_t^{(u)}),\tilde{\psi}_t^{(u)}).
\end{eqnarray}

At $\Phi\rightarrow 0$, Eq. (39) yields the following expression for
the minimal (along the boundary) deviation of energy from the
barrier:
\begin{eqnarray}
\delta_{\min}^{(u)}&\equiv&
E_b^{(2)}-E_{\max}^{(u)}=(E_b^{(2)}-E_t^{(u)})/\sqrt{{\rm e}}=
\nonumber
\\
&=&\frac{\pi^{3/2}} {2{\rm
e}^{1/2}}\frac{h}{\sqrt{\ln\left(1/\Phi\right)/n}}.
\end{eqnarray}

\vskip 0.2cm

\hskip 2.9cm {\it b. Relatively large $h$}

\vskip 0.2cm

If $h>h_{cr}^{(u)}(\omega_f)$ (cf. Fig. 8(a)), the previously described
trajectory (tangent one or the separatrix, for the even and odd
spikes ranges respectively) is encompassed by the separatrix of the
upper nonlinear resonance and typically forms the boundary of the
major stability island inside the upper layer (reproduced
periodically in $\tilde{\psi}$ with the period $2\pi$). The lower
{\it outer} boundary of the layer is formed in this case by the
lower part of the {\it resonance separatrix}. This may be
interpreted as the absorption of the upper resonance by the upper
chaotic layer.

\vskip 0.5cm

The description of chaotic layers given above and, in more details,
in the Appendix is the {\bf first main result of this paper}. It
provides a {\it rigorous base} for our intuitive assumption that the
minimal value of $h$ at which the layers overlap corresponds to the
reconnection of the nonlinear resonances with each other while the
reconnected resonances touch one of the layers and touch/overlap
another layer. It is remarkable also that we have managed to obtain
the {\it quantitative} theoretical description of the chaotic layers
boundaries in the {\it phase space}, including even the major
stability islands, that well fits the results of simulations (see
Fig. 8(b)).

\subsection{Onset of global chaos: the spikes minima}

The condition for the merger of the lower resonance and the lower
chaotic layer may be written as

\begin{equation}
\tilde{H}(I=I(E=E_b^{(1)}+\delta^{(l)}_{\min}),\tilde{\psi}=0)=\tilde{H}_{sl}.
\end{equation}

The condition for the merger of the upper resonance and the upper
chaotic layer may be written as

\begin{equation}
\tilde{H}(I=I(E=E_b^{(2)}-\delta^{(u)}_{\min}),\tilde{\psi}=\pi)=\tilde{H}_{su}.
\end{equation}

For the global chaos onset related to the spike minimum, either of
Eqs. (41) and (42) should be combined with the condition of the
separatrix reconnection (20). Let us seek first only the leading
terms of $h_s\equiv h_s(\Phi)$ and $\omega_s\equiv \omega_s(\Phi)$.
Then (20) may be replaced by its lowest-order approximation (21) or,
equivalently, (22). Using also the lowest-order approximation for
the barriers ($E_b^{(1,2)}\approx 1/2\mp \Phi$), we reduce Eqs.
(41), (42) respectively to

\begin{eqnarray}
&&
\tilde{H}(I=I(E=1/2-\Phi+\delta^{(l)}_{\min}),\tilde{\psi}=0)=0,
\\
&&
\tilde{H}(I=I(E=1/2+\Phi-\delta^{(u)}_{\min}),\tilde{\psi}=\pi)=0,
\end{eqnarray}

\noindent where $\delta^{(l)}_{\min}$ is given by (28) or (31) for
the odd or even spikes respectively while $\delta^{(u)}_{\min}$ is
given by (35) or (40) for the odd or even spikes respectively.

The solution $(h_s^{(l)},\omega_s^{(l)})$ of the system of equations
(22),(43) and the solution $(h_s^{(u)},\omega_s^{(u)})$ of the
system of equations (22),(44) turn out {\it identical} to the {\it
leading} order. For the sake of brevity, we derive below just
$(h_s^{(l)},\omega_s^{(l)})$, denoting the latter, in short, as
$(h_s,\omega_s)$ \cite{28_prime}.

The system of algebraic equations (22) and (43) is still too
complicated to find its exact solution. However, we need only the
{\it lowest-order} solution - and this simplifies the problem.
Still, even this simplified problem is not trivial, both because the
function $h_s(\Phi)$ turns out to be non-analytic and because
$\Delta E^{(1)}$ in (22) is very sensitive to $\omega_f$ in the
relevant range. Despite these difficulties, we have found the
solution in a {\it self-consistent} way, as briefly described below.

Assume that the asymptotic dependence $h_s(\Phi) $ is:

\begin{eqnarray}
&&
h_s=a\frac{\Phi}{\ln(4{\rm e}/\Phi)} \, ,
\end{eqnarray}

\noindent where the constant $a$ may be found from the
asymptotic solution of (22), (43), (45).

Substituting the energies $E = 1/2-\Phi+\delta^{(l)}_{\min}$ and $
E =1/2+ \Phi- \delta^{(u)}_{\max}$ in (7) and taking
into account (28), (31), (35), (40) and (45), we find that, both for
the odd and even spikes, the inequality

\begin{equation}
\omega_m-\omega(E)\ll \omega_m
\end{equation}

\noindent
holds in the whole relevant range of energies, i.e. for

\begin{equation}
\Delta E\in [-\Phi+\delta^{(l)}_{\min},\Phi-\delta^{(u)}_{\min}].
\end{equation}

\noindent Thus, the resonant approximation is valid in the whole
range (47). Eq. (12) for $q_n(E)$ is valid in the whole
relevant range of energies too.

Consider Eq. (43) in a more explicit form. Namely, we express
$\omega_f$ from (43), using Eqs. (4), (12), and (13), and using also
(28)/(31) for odd/even spikes, and (45):

\begin{equation}
\omega_f =  \frac{n\pi}{2 \ln \left(\frac{4{\rm e}}{\Phi} \right)
} \left\{ 1+\frac{h\sqrt{2}}{n\Phi} + O \left(
\frac{1}{\ln^2(4{\rm e}/\Phi)}\right) \right\}.
\end{equation}

\noindent We emphasize that the value of $\delta_{\min}^{(l)}$
enters explicitly only the term $O(\ldots)$ while, as it is clear
from the consideration below, this term does not affect the leading
terms in $(h_s,\omega_s)$. Thus, $\delta_{\min}^{(l)}$ does not
affect the leading term of $\omega_s$ at all, while it affects the
leading term of $h_s$ only {\it implicitly}: $\delta_{\min}^{(l)}$
lies in the range of energies where $nq_n(E)\approx \sqrt{2}$. This
latter quantity is present in the second term in the curly brackets
in (48) and, as it is clear from the further consideration, $h_s$ is
proportional to it.

Substituting (48) into the expression for $\Delta E^{(1)}$ in
(22), using (45) and keeping only the leading terms, we obtain

\begin{equation}
\Delta E^{(1)}=\Phi\sqrt{1-4{\rm e}^{c-2}}, \quad\quad c\equiv
\frac{2\sqrt{2}}{n}a.
\end{equation}

Substituting $\Delta E^{(1)}$ from (49) into Eq. (22) for
$h(\omega_f)$ and allowing for (45) once again, we arrive at the
transcendental equation for $c$:

\begin{eqnarray}
&&
\ln
\left(
\frac{1+\chi(c)}{1-\chi(c)}
\right)
-2\chi(c)=c,
\\
&&
\chi(c)\equiv \sqrt{1-4{\rm e}^{c-2}}.
\nonumber
\end{eqnarray}

\noindent The approximate numerical solution of Eq. (50) is:

\begin{equation}
c \simeq0.179.
\end{equation}

Thus, the final leading-order asymptotic formulas for $\omega_f$
and $h$ in the minima of the spikes are the following:

\begin{eqnarray}
&&
\omega_{s0}\equiv \omega_{s0}^{
\left(\frac{n+1}{2}
\right)}=n\frac{\pi}{2\ln
\left(\frac{4{\rm e}}{\Phi}
\right) },
\\
&&
h_{s0}\equiv h_{s0}^{\left(\frac{n+1}{2}\right)}=n\frac{c}{2\sqrt{2}}\frac{\Phi}{\ln
\left(\frac{4{\rm e}}{\Phi}
\right) },
\nonumber
\\
&& n=1,3,5,..., \quad\quad \Phi\rightarrow 0, \nonumber
\end{eqnarray}

\noindent where the constant $c\simeq 0.179$ is the solution of
Eq. (50).

The rigorous derivation of the explicit asymptotic formulas for the
minima of $h_{gc}(\omega_f)$ is the {\bf second main result of this
paper}. These formulas allow one to immediately predict the
parameters for the weakest perturbation which may lead to global
chaos.

\subsection{Numerical example and next-order corrections}

For $\Phi=0.2$, the numerical simulations give the following
values for the frequencies in the minima of the first two spikes
(see Fig. 4):

\begin{equation}
\omega_{s}^{(1)}\approx 0.4005\pm 0.0005, \quad\quad
\omega_{s}^{(2)}\approx 1.24\pm 0.005.
\end{equation}

The values by the lowest-order formula (52) are:

\begin{equation}
\omega_{s0}^{(1)}\approx 0.393, \quad\quad \omega_{s0}^{(2)}\approx
1.18,
\end{equation}

\noindent in rather good agreement with the simulations.

The next-order correction for $\omega_s$ can be immediately found
from Eq. (48) for $\omega_f$ and Eq. (52) for $h_{s0}$, so that

\begin{eqnarray}
&& \omega_{s1}\simeq\omega_{s0}(1+\frac{c}{2\ln \left(\frac{4{\rm
e}}{\Phi} \right)})\approx \frac{n\pi \left( 1+\frac{0.09}{\ln
\left(\frac{4{\rm e}}{\Phi} \right) } \right) }{2\ln
\left(\frac{4{\rm e}}{\Phi} \right) } \, ,
\\
&& n=1,3,5,... \nonumber
\end{eqnarray}

The formula (55) agrees with the simulations even better than the
lowest-order approximation:

\begin{equation}
\omega_{s1}^{(1)}\approx 0.402, \quad\quad
\omega_{s1}^{(2)}\approx 1.21.
\end{equation}

For $h$ in the spikes minima, the simulations give the following
\cite{26_prime} values (see Fig. 4):

\begin{equation}
h_{s}^{(1)}\approx 0.0049, \quad\quad h_{s}^{(2)}\approx 0.03.
\end{equation}

The values by the lowest-order formula (52) are:

\begin{equation}
h_{s0}^{(1)}\approx 0.0032, \quad\quad h_{s0}^{(2)}\approx 0.01.
\end{equation}

The theoretical value $h_{s0}^{(1)}$ gives a reasonable estimate for
the simulation value $h_{s}^{(1)}$. The theoretical value
$h_{s0}^{(2)}$ gives the correct order of magnitude for the
simulation value $h_{s}^{(2)}$. Thus, the accuracy of the
lowest-order formula (52) for $h_s$ is much lower than that for $
\omega_s $: this is due to the steepness of $h_{gc}(\omega_f)$ in
the ranges of spikes (the steepness, in turn, is due to the flatness
of the function $\omega(E)$ near its maximum). Moreover, as the
number of the spike $j$ increases, the accuracy of the lowest-order
value $h_{s0}^{(j)}$ significantly decreases. The latter can be
explained as follows. For the next-order correction to
$h_{s0}^{(j)}$, the dependence on $\Phi$ reads as:

\begin{equation}
\frac{h_{s1}^{(j)}-h_{s0}^{(j)}}{h_{s0}^{(j)}}\propto \frac{1}{\ln(4{\rm
e}/\Phi)}.
\end{equation}

\noindent At least some of the terms of this correction are positive
and proportional to $h_{s0}^{(j)}$ (e.g. due to the difference
between the exact equation (15) and its approximate version (17))
while $h_{s0}^{(j)}$ is proportional to $n\equiv 2j-1$. Thus, for
$\Phi=0.2$, the relative correction for the 1st spike is $\sim 0.25$
while the correction for the 2nd spike is a few times larger i.e.
$\sim 1$. But the latter means that, for $\Phi=0.2$, the asymptotic
theory for the 2nd spike cannot pretend to be a quantitative
description of $h_s^{(2)}$, but only provide the correct order of
magnitude. Besides, if $n>1$ while $\Phi$ exceeds some critical
value, then the search of the minimum involves Eq. (66) rather than
Eq. (20), as explained below in Sec. IV (cf. Figs. 10(b) and 11).
Altogether, this explains why $h_{s}^{(1)}$ is larger than
$h_{s0}^{(1)}$ only by 50$\%$ while $h_{s}^{(2)}$ is larger than
$h_{s0}^{(2)}$ by 200$\%$.

The consistent explicit derivation of the correction to
$h_{s0}^{(j)}$ is complicated. A reasonable alternative may be a
proper {\it numerical} solution of the algebraic system of Eqs. (20)
\cite{32_prime} and (41) for the odd spikes or (42) for the even
spikes \cite{28_prime,29_prime}. To this end, in Eqs. (20)
\cite{32_prime} and (41)/(42) we use: (i) the exact values of the
saddle energies obtained from the exact relations (16) instead of
the approximate relations (17); (ii) the exact formulas (5) and (6)
for $\omega(E)$ instead of the asymptotic expression (7); (iii) the
exact functions $q_n(E)$ instead of the asymptotic formula (12);
(iv) the relation (27) and the calculation of the \lq\lq tangent''
state $(\tilde{\psi}_t^{(l)}, I_t^{(l)})$ by Eqs. (A11), (A22) for
the odd spikes, or relation (39) and the calculation of the \lq\lq
tangent'' state $(\tilde{\psi}_t^{(u)}, I_t^{(u)})$ by Eqs.
(A41)-(A43) for the even spikes. Note that, to find the exact
function $q_n(E)$, we substitute into the definition of $q_n(E)$ in
(4) the explicit \cite{general_case} solution for $q(E,\psi)$:

\begin{eqnarray}
&& q(E,\psi)=\arcsin \left( \frac{\eta-\sqrt{2E}+\Phi}{1-\eta}
\right) \quad {\rm for} \quad \psi\in\left[0,\frac{\pi}{2}\right],
\nonumber
\\
&& q(E,\psi)=\pi-q(E,\pi-\psi) \quad\quad\quad\quad {\rm for}
\quad \psi\in\left[\frac{\pi}{2},\pi\right], \nonumber
\\
&& q(E,\psi)=q(E,2\pi-\psi) \quad\quad\quad\quad\quad\quad {\rm
for} \quad \psi\in\left[\pi,2\pi\right], \nonumber
\\
&& \eta\equiv \frac{1}{2}(\sqrt{2E}-\Phi+1){\rm sn}^2
\left(\frac{2K}{\pi}\psi \right) \, ,
\end{eqnarray}
where ${\rm sn}(x)$ is the elliptic sine \cite{Abramovitz_Stegun}
with the same modulus $k$ as the full elliptic integral $K$
defined in (5),(6).

%For the first spike,
The numerical solution described above
gives:

%\begin{equation}
\begin{eqnarray}
&&   \left(\omega_{s}^{(1)}\right)_{num}\approx 0.401 \ , \qquad
   \left(h_{s}^{(1)}\right)_{num} \approx 0.005 \ ,
   \nonumber
\\
&&\\
 &&   \left(\omega_{s}^{(2)}\right)_{num}\approx 1.24 \ , \qquad
   \left(h_{s}^{(2)}\right)_{num} \approx 0.052 \ .
\nonumber
\end{eqnarray}
%\end{equation}

The agreement with the simulation results is: (i) excellent for
$\omega_{s}$ for the both spikes and for $h_s$ for the 1st spike,
(ii) reasonable for $h_s$ for the 2nd spike. Thus, if $\Phi$ is {\it
moderately} small, a much more accurate prediction for $h_s$ than
that by the lowest-order formula is provided by the numerical
procedure described above.

\section{THEORY OF THE SPIKES WINGS}

The goal of this section is
to
%understand the
find mechanisms responsible for the formation of the spikes {\it
wings} (i.e. the function $h_{gc}(\omega_f)$ in the ranges of
$\omega_f$ slightly deviating from $\omega_s^{(j)}$) and to provide
for their theoretical description.
% of the wings.
%This is the main aim of this section.

Before developing the theory, we briefly analyze the simulation data
(Fig. 4), concentrating on the 1st spike. The left wing of the spike
is smooth and nearly straight. The initial part of the right wing is
also nearly straight \cite{fluctuations} though less steep. But, at
some small distance from $\omega_s^{(1)}$, its slope changes
jump-wise by a few times: compare the derivative \cite{fluctuations}
${\rm d}h_{gc}/{\rm d}\omega_f\approx 0.1$ at $\omega_f$ slightly
exceeding $\omega_s^{(1)}\approx 0.4$ (see the left inset in Fig. 4)
and ${\rm d}h_{gc}/{\rm d}\omega_f\approx 0.4$ at $\omega_f = 0.45
\div 0.55$ (see the main part of Fig. 4). Thus, even prior to the
theoretical analysis, one may assume that there are a few different
important mechanisms responsible for the formation of the wings.

Consider the arbitrary $j$th spike. We have shown in the previous
section that, in the asymptotic limit $\Phi\rightarrow 0$, the
minimum of the spike corresponds to the intersection between the
lines (20) and (41)/(42) for odd/even spikes. We recall that: (i)
Eq. (20) corresponds to the overlap in phase space between nonlinear
resonances of the same order $n\equiv 2j-1$; (ii) Eq. (41)/(42)
corresponds to the onset of the overlap between the resonance
separatrix associated with the lower/upper saddle and the chaotic
layer associated with the lower/upper potential barrier; (iii) for
$\omega_f=\omega_s^{(j)}$, the condition (41)/(42) guarantees also
the overlap between the upper/lower resonance separatrix and the
chaotic layer associated with the upper/lower barrier
\cite{28_prime}.

\begin{figure}[tb]
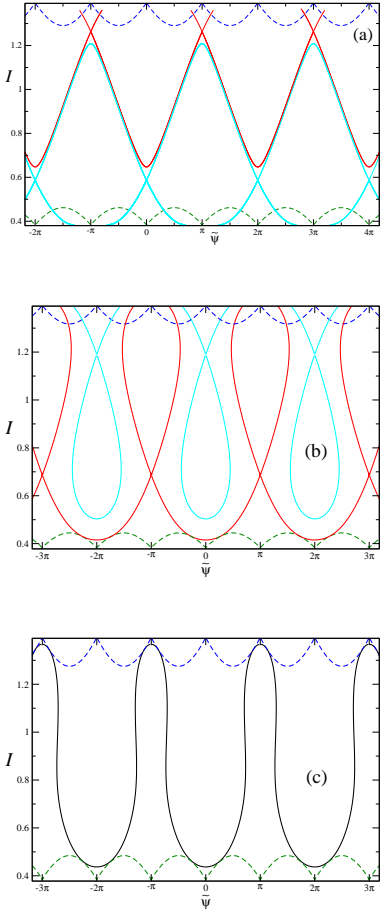

\includegraphics[width = 5. cm]{prefig9a.eps}
\vskip 0.75 cm
\includegraphics[width = 5. cm]{prefig9b.eps}
\vskip 0.75 cm
\includegraphics[width = 5. cm]{prefig9c.eps}
\vskip 0.75 cm \caption{Illustrations of the mechanisms of the
formation of the 1st spike wings and of the corresponding
theoretical lines in Fig. 10(a). Boxes (a), (b) and (c) illustrate
the lines by Eqs. (20), (41) and (64) respectively: the
corresponding perturbation parameters are
($\omega_f=0.39,h=0.0077$), ($\omega_f=0.41,h=0.00598$) and
($\omega_f=0.43,h=0.01009$) respectively. Resonance separatrices
are drawn by red and cyan. The dashed lines show the functions
$I_{\rm GSS}^{(l)}(\tilde{\psi})$ and $I_{\rm
GSS}^{(u)}(\tilde{\psi})$. The black line in (c) is the trajectory
of the resonant Hamiltonian system (4) which is tangent to both
dashed lines.}
\end{figure}

If $\omega_f$ becomes slightly smaller than $\omega_s^{(j)}$ the
resonances shift closer to the barriers while moving apart from each
other. Hence, as $h$ increases, the overlap of the resonances with
the chaotic layers associated with the barriers occurs earlier than
with each other. Therefore, at $0<\omega_s^{(j)}- \omega_f\ll
\omega_m$, the function $h_{gc}(\omega_f)$ should approximately
correspond to the reconnection of resonances of the order $n\equiv
2j-1$ (Fig. 9(a)). Fig. 10(a) demonstrates that even the asymptotic
formula (22) for the separatrix reconnection line fits the left wing
of the 1st spike quite well while the numerically calculated line
(20) agrees with the simulations perfectly.

If $\omega_f$ becomes slightly larger than $\omega_s^{(j)}$ then, on
the contrary, the resonances shift closer to each other and more far
from the barriers. Therefore, the overlap of resonances with each
other occurs at smaller $h$ than the overlap between any of them and
the chaotic layer associated with the lower/upper barrier (cf. Figs.
5(c) and 5(d) as well as 6(c) and 6(d)). Hence, it is the latter
overlap which determines the function $h_{gc}(\omega_f)$ in the relevant
range of $\omega_f$ (Fig. 9(b)). Fig.~10 shows that
$h_{gc}(\omega_f)$ is indeed well approximated in the close vicinity
to the right from $\omega_s^{(j)}$ by the numerical solution of Eq.
(41)/(42) for an odd/even spike and, for the 1st spike and the given
$\Phi$, even by its asymptotic form,

\begin{eqnarray}
&& h\equiv h(\omega_f)=
\\
\nonumber
&&
n\frac
{-\Phi +\frac{\omega_f}{n\pi}
\left[
\Phi
\left\{
2\ln\left(\frac{4{\rm e}}{\Phi}\right)+\ln\left(\frac{\Phi+\Delta E^{(1)}}{\Phi-\Delta E^{(1)}} \right)
\right\}
-2\Delta E^{(1)}
\right]
}
{2\sqrt{2}},
\\
&& \Delta E^{(1)}\equiv
\sqrt{\Phi^2-64\exp(-\frac{n\pi}{\omega_f})}, \nonumber \\
&& n\equiv 2j-1,\quad |\omega_f-\omega_s^{(j)}|\ll \omega_m \, ,
\nonumber
\end{eqnarray}

\begin{figure}[tb]
\includegraphics[width = 6.8 cm]{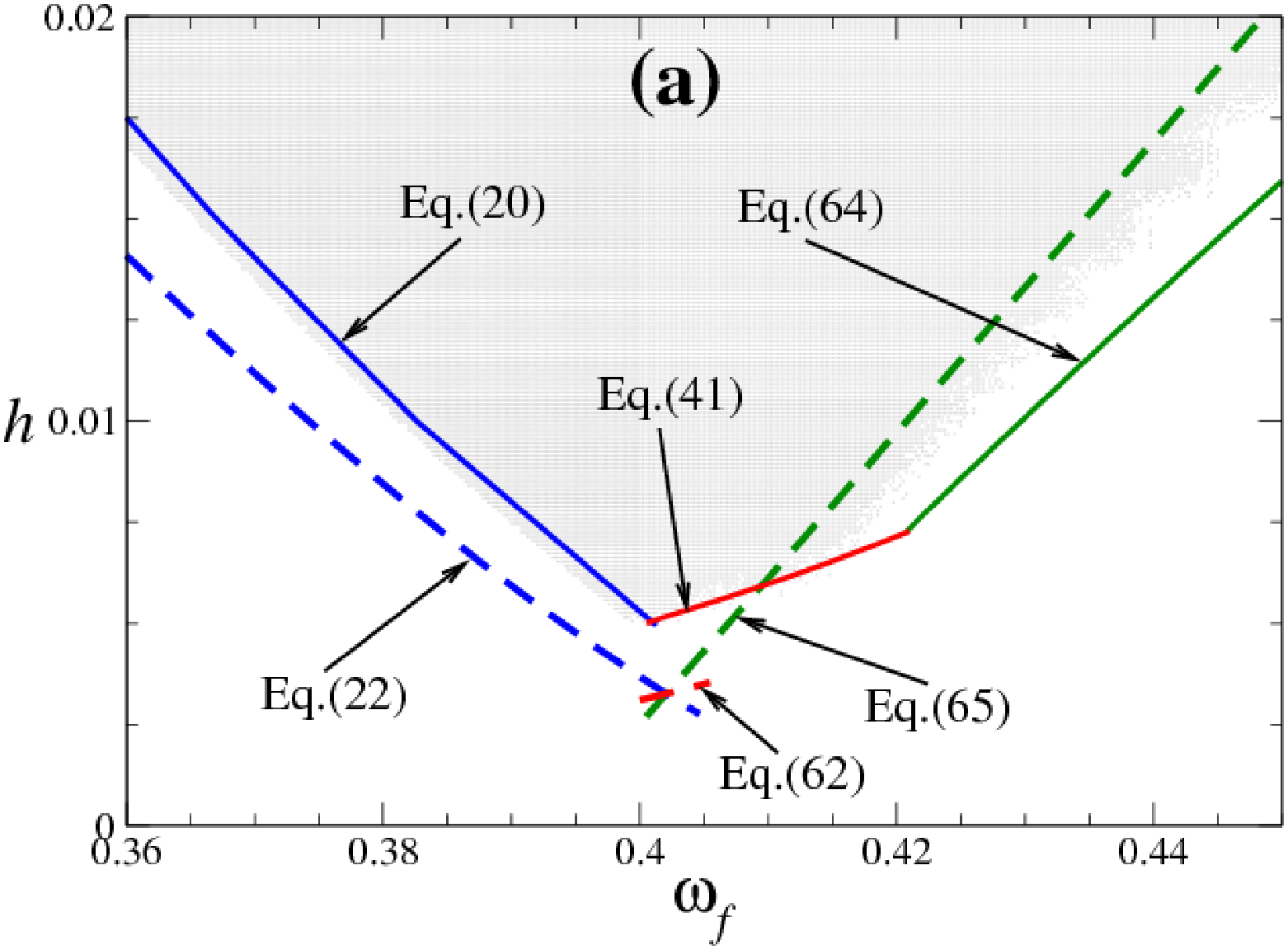}
\vskip 0.75 cm
\includegraphics[width = 6.8 cm]{prefig10b.eps}
\caption{The 1st (a) and 2nd (b) spike in $h_{gc}(\omega_f)$:
comparison between the results of the numerical simulations (the
lower boundary of the shaded area) and the theoretical estimates.
The estimates are indicated by the corresponding equation numbers
and are drawn by different types of lines, in particular the dashed
lines represent the explicit asymptote for the solid line of the same
color.}
\end{figure}

The mechanism described above determines $h_{gc}(\omega_f)$ only in
the close vicinity of $\omega_s^{(j)}$. If $\omega_f/n$ becomes too
close to $\omega_m$ or exceeds it, then the resonances are not of
immediate relevance: they may even disappear or, if they still
exist, their closed loops shrink, so that they cannot anymore
provide for the connection of the chaotic layers in the relevant
range of $h$. At the same time, the closeness of the frequency to
$\omega_m$ still may give rise to a large variation of action along
the trajectory of the Hamiltonian system (4). For the odd/even spikes,
the boundaries of the chaotic layers in the asymptotic limit
$\Phi\rightarrow 0$ are formed in this case by the trajectory of (4)
which is tangent to the lower/upper GSS curves (for the lower/upper
layer) or by the lower/upper part of the separatrix of (4) generated
by the saddle \lq\lq $\tilde{s}$''/\lq\lq $s$'' (for the upper/lower
layer). Obviously, the overlap of the layers occurs when these
trajectories coincide with each other, that may be formulated as the
equality of $\tilde{H}$ in the corresponding tangency and saddle:

\begin{eqnarray}
&& \tilde{H}(I_t^{(l)},\tilde{\psi}_t^{(l)})=
\tilde{H}(I_{\tilde{s}},\tilde{\psi}_{\tilde{s}}) \quad
{\rm for}
\quad j=1,3,5,\ldots, \nonumber
\\
&&\tilde{H}(I_s,\tilde{\psi}_s)=
\tilde{H}(I_t^{(u)},\tilde{\psi}_t^{(u)}) \quad {\rm for} \quad
j=2,4,6,\ldots, \nonumber
\\
&& I_{\tilde{s}}\equiv I(E_b^{(2)}-\delta_{\tilde{s}}), \quad\quad
I_{s}\equiv I(E_b^{(1)}+\delta_{s}).
\end{eqnarray}

\noindent Note however that, for {\it moderately} small $\Phi$,
the tangencies may be relevant both to the lower layer and to the
upper  one (see the Appendix). Indeed, such a case occurs for our
example with $\Phi=0.2$: see Fig. 9(c). Therefore,
the overlap of the layers corresponds to the equality of
$\tilde{H}$ in the tangencies:

\begin{equation}
\tilde{H}(I_t^{(l)},\tilde{\psi}_t^{(l)})=
\tilde{H}(I_t^{(u)},\tilde{\psi}_t^{(u)})
\, .
\end{equation}

To the lowest order,
% asymptotic form of both
Eq. (63) and Eq. (64) read as:

\begin{equation}
h\equiv h(\omega_f)=\frac{\sqrt{2}\Phi\ln\left(\frac{4{\rm
e}}{\Phi}\right)}
{\pi}\left(\omega_f-\frac{n\pi}{2\ln\left(\frac{4{\rm
e}}{\Phi}\right)}
 \right).
\end{equation}

Both the line (64) and the asymptotic line (65) well agree with the
part of the right wing of the 1st spike situated beyond the
immediate vicinity of the minimum from the right side, namely, to
the right from the fold at $\omega_f\approx 0.42$ (Fig. 10(a)). The
fold corresponds to the change of the mechanisms of the chaotic
layers overlap.

\begin{figure}[tb]
\includegraphics[width = 5. cm]{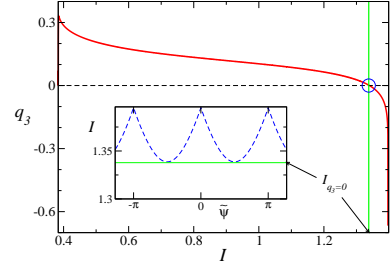}
\caption{Amplitude of the 3rd Fourier harmonic as a function of
action (solid red line). The dashed black line shows the zero level.
Its intersection with the solid red line is marked by the circle.
The green line indicates the value of action where $q_3=0$. The
inset illustrates the line (66) in Fig. 10(b): the GSS curve touches
the horizontal line $I=I_{q_3=0}$.}
\end{figure}

If $\Phi$ is moderately small while $n>1$, the description of the
far wings by the numerical lines (20) and (64) may be still quite
good but the asymptotic lines (22) and (65) cannot pretend to
describe the wings quantitatively anymore (Fig. 10(b)). As for the
very minimum of the spike and the wings in the close vicinity to it,
one more mechanism may become relevant for their formation in this
case (Figs. 10(b) and 11). This mechanism may be explained as
follows. If $n>1$, then $q_n(E)$ becomes zero in the close vicinity
($\sim \Phi^2$) of the relevant barrier (the upper/lower barrier, in
the case of even/odd spikes: cf. Fig. 11). As follows from the
equations of motion (14), no trajectory can cross the line
$I=I_{q_n=0}$. In the asymptotic limit $\Phi\rightarrow 0$, provided
$h$ is from the relevant range, almost the whole GSS curve is farer
from the barrier than the line $I=I_{q_n=0}$, and the latter becomes
irrelevant. But, for a moderately small $\Phi$, the line may
separate the whole GSS curve from the rest of the phase space. Then
the resonance separatrix cannot connect to the GSS curve even if
there is a state on the latter curve with the same value of
$\tilde{H}$ as on the resonance separatrix.
%In this case,
For a given $\omega_f$, the connection requires then a higher value
of $h$: for such a value, the GSS curve itself crosses the line
$I=I_{q_n=0}$.
%Given that,
In the relevant range of $h$, the resonance separatrix passes very
close to this line, so that the connection is well approximated by
the condition that the GSS curve {\it touches} this line (see the
inset in Fig. 11):

\begin{eqnarray}
&& \delta_u=E_b^{(2)}-E_{q_{2j-1}=0}\quad\quad {\rm for}\quad
j=2,4,6,\ldots,
\nonumber\\
&&\\
&& \delta_l=E_{q_{2j-1}=0}-E_b^{(1)}\quad\quad {\rm for}\quad
j=3,5,7,\ldots, \nonumber
\end{eqnarray}

\noindent This mechanism is relevant for the formation of the
minimum of the 2nd spike at $\Phi=0.2$, and in the close vicinity of the
spike, on the left (Fig. 10(b)).

Finally, let us explicitly find the {\it universal asymptotic shape}
of the spike in the vicinity of its minimum.

First, we note that the lowest-order expression (62) for the spike
between the minimum and the fold can be written as the {\it
half-sum} of the expressions (22) and (65) (which represent the
lowest-order approximations for the spike to the left of the
minimum, and to the right of the fold respectively). Thus, all three
lines (22), (62) and (65) intersect in one point. The latter means
that, in the asymptotic limit $\Phi\rightarrow 0$, the fold merges
with the minimum: $\omega_f$ and $h$ in the fold asymptotically
approach $\omega_s$ and $h_s$ respectively. Thus, though the fold is
a generic feature of the spikes, it is not of main significance: the
spike is formed basically by two straight lines. The ratio between
their slopes is universal. So, introducing a proper scaling, we
reduce the spike shape to a universal function (Fig. 12):

\begin{eqnarray}
&& \tilde{h}(\Delta \tilde {\omega}_f)=\tilde{h}^{(lw)}(\Delta
\tilde {\omega}_f)\equiv 1-\sqrt {1-4{\rm e}^{c-2}}\Delta \tilde
{\omega}_f\approx \nonumber\\
&& \quad\quad\quad\quad\quad\quad\quad\quad\approx 1-0.593\Delta
\tilde {\omega}_f \quad {\rm for}\quad \Delta \tilde {\omega}_f<0,
\nonumber
\\
&& \tilde {h}(\Delta \tilde {\omega}_f) = \tilde{h}^{(rw)}(\Delta
\tilde {\omega}_f)\equiv 1+\Delta \tilde
 {\omega}_f\quad {\rm for}\quad\Delta \tilde {\omega}_f>0, \nonumber
\\
&&  \\
&&
 \tilde{h}^{(fold)}(\Delta
\tilde {\omega}_f)= \frac{ \tilde{h}^{(lw)}(\Delta \tilde
{\omega}_f)+ \tilde{h}^{(rw)}(\Delta
\tilde {\omega}_f)}{2}\equiv\nonumber \\
&&\equiv
 1+\frac{1-\sqrt {1-4{\rm
e}^{c-2}}}{2}\Delta \tilde {\omega}_f \approx 1+0.203\Delta \tilde
{\omega}_f, \nonumber\\
&& \nonumber \\
&& \tilde {h}\equiv \frac{h}{h_{s0}},  \quad\quad\Delta \tilde
{\omega}_f\equiv
\frac{\omega_f-\omega_{s1}}{\omega_{s1}-\omega_{s0}}, \quad\quad
\Phi\rightarrow 0,\nonumber
\end{eqnarray}

\noindent where $\omega_{s0}$ and $h_{s0}$ are the lowest-order
expressions (52) respectively for the frequency and amplitude in the
spike minimum, $\omega_{s1}$ is the expression (55) for the
frequency in the spike minimum, including the first-order
correction, and $c$ is the constant (51).

Beside the left and right wings of the universal shape (the solid
lines in Fig. 12), we also present in (67) the function
$\tilde{h}^{(fold)}(\Delta \tilde {\omega}_f)$ (the dashed line in
Fig. 12): it purposes to show that, on one hand, the fold
asymptotically merges with the minimum but, on the other hand, the
fold is generic and the slope of the spike between the minimum and
the fold has a universal ratio to any of the slopes of the major
wings.

Even for a moderately small $\Phi$, like in our example, the ratios
between the three slopes related to the 1st spike in the simulations are
reasonably well reproduced by those in Eq. (67): cf. Figs. 10(a) and
12. It follows from (67) that the asymptotic scaled shape is
universal i.e. independent of $\Phi$, $n$ or any other parameter.

\begin{figure}[tb]
\includegraphics[width = 5. cm]{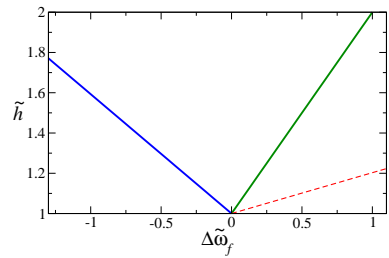}
\caption{The universal shape of the spike minimum (67) (solid
lines). The dashed line indicates the universal slope of the spike
in between the minimum and the fold, which have merged in the
universal (asymptotic) function (67).}
\end{figure}

The description of the wings of the spikes near their minima, in
particular the derivation of the spike universal shape, is the {\bf
third main result of this paper}.

\section{GENERALIZATIONS AND APPLICATIONS}

The {\it new approach} for the treatment of separatrix chaos opens a
broad variety of important generalizations and applications, some of
which are discussed below.

\begin{enumerate}

\item[{\bf 1.}] It may be applied to {\it any} separatrix layer,
including in particular the conventional {\it single-separatrix}
case. This is possible due to the characteristic dependence of the
frequency of eigenoscillation on energy in the vicinity of any
separatrix: the frequency keeps nearly a {\it constant} value even
if the deviation of the energy from the separatrix {\it strongly
varies} within a given scale of the deviation. To the best of our
knowledge, the latter idea was not exploited before, which is why
probably it was thought to be impossible to analyze the dynamics of
the separatrix map in an explicit form. There were various estimates
of the layer width in energy (see
%books and reviews
\cite{Chirikov:79,lichtenberg_lieberman,Zaslavsky:1991,zaslavsky:1998,zaslavsky:2005,treshev}
and references therein) but the quantitative analysis of the layer
boundaries in the {\it phase space} was never done in the
non-adiabatic case. In contrast, our approach does analyze the
dynamics of the separatrix map, and this allows us in particular to
quantitatively describe the layer boundaries in case when the
perturbation is resonant with the eigenoscillation in the relevant
range of energy. It follows from such a description that, for a
given small amplitude $h$ of the perturbation, the maximal layer
width in energy is much larger than it is assumed by former theories
(cf.
\cite{Chirikov:79,lichtenberg_lieberman,Zaslavsky:1991,zaslavsky:1998,zaslavsky:2005}
where the maximal width is assumed to be $\sim h$). Thus, we can
quantitatively describe large {\it jumps} and {\it peaks} of the
layer width
%in energy
as a function of the perturbation amplitude
and frequency respectively \cite{soskin2000,pr}. Moreover, a rough
estimate on the basis of our approach indicates that, for many
classes of systems, the relative range of such jumps/peaks (i.e. the
ratio between the upper and lower levels of the jump/peak) {\it
diverges} in the asymptotic limit $h\rightarrow 0$.
%of a vanishing perturbation amplitude.

\item[{\bf 2.}] Apart from the description of the boundaries, the
approach allows us to describe the {\it chaotic transport} within
the layer. In particular, it may allow us to calculate a positive
Lyapunov exponent and to describe diffusion.

\item[{\bf 3.}] Our approach may be generalized
for a {\it non-resonant} perturbation. The resonance approximation
is not valid then but there still remains the property of a nearly
constancy of the frequency of eigenoscillation within an arbitrary
given scale of the deviation of energy from the separatrix. This
property may allow us to explicitly describe the dynamics of the
separatrix map for any frequency of perturbation which is less than
or of the order of the resonance frequency.
%, and possibly even for larger frequencies.

\item[{\bf 4.}] Apart from Hamiltonian systems of the one and a half degrees
of freedom and corresponding Zaslavsky separatrix maps, our approach
may be useful in the treatment of other chaotic systems and
separatrix maps (see \cite{treshev} for the most recent major review
on various types of separatrix maps and related continuous chaotic
systems).
\end{enumerate}

As concerns the {\it facilitation of the global chaos onset} between
adjacent separatrices, we mention the following generalizations:

{\bf 1.} The spikes in $h_{gc}(\omega_f)$ may occur for an {\it
arbitrary Hamiltonian} system with two or more separatrices. The
asymptotic theory can be generalized accordingly.

{\bf 2.} The absence of pronounced spikes at {\it even} harmonics
$2j\omega_m$ is explained by the symmetry of the potential (1): the
even Fourier harmonics of the coordinate, $q_{2j}$, are equal to
zero. For time-periodic perturbation of dipole type, like in Eq.
(2), there are no resonances of even order due to this symmetry
\cite{Chirikov:79,lichtenberg_lieberman,Zaslavsky:1991,zaslavsky:1998,zaslavsky:2005,pr}.
If either the potential is {\it non-symmetric} or the additive
perturbation of the Hamiltonian is not an {\it odd} function of the
coordinate, then even-order resonances do exist, resulting in the
presence of the spikes in $h_{gc}(\omega_f)$ at $\omega_f \approx
2j\omega_m$.

{\bf 3.} There may be an additional facilitation of global chaos
onset which is reasonable to call a \lq\lq secondary'' facilitation.
Let the frequency $\omega_f$ be close to the frequency $\omega_s$ of
the spike minimum while the amplitude $h$ be $\sim h_s$ but still
lower than $h_{gc}(\omega_f)$. Then there are two resonance
separatrices in the $I-\tilde{\psi}$ plane which are not connected
by the chaotic transport (cf. Fig. 6(b) and 5(b)). This system
possesses the zero-dispersion property. The trajectories of the
resonant Hamiltonian (4) which start in between the separatrices
oscillate in $I$ (as well as in ${\rm d}\tilde{\psi}/{\rm d}t$). The
frequency $\tilde{\omega}$ of such oscillations along a given
trajectory depends on the corresponding value of $\tilde{H}$
analogously to as $\omega$ depends on $E$ for the original
Hamiltonian $H_0$: $\tilde{\omega}(\tilde{H})$ is equal to zero for
the values of $\tilde{H}$ corresponding to the separatrices (being
equal in turn to $\tilde{H}_{sl}$ and $\tilde{H}_{su}$: see Eq.
(19)) while possessing a nearly rectangular shape in between,
provided the quantity $|\tilde{H}_{sl}-\tilde{H}_{su}|$ is much
smaller than the variation of $\tilde{H}$ within each of the
resonances,

\begin{equation}
|\tilde{H}_{sl}-\tilde{H}_{su}| \ll \tilde{H}_{var}\sim
|\tilde{H}_{sl}-\tilde{H}_{el}|\sim
|\tilde{H}_{su}-\tilde{H}_{eu}|,
\end{equation}

\noindent where $\tilde{H}_{el}$ and $\tilde{H}_{eu}$ are the values
of $\tilde{H}$ in the elliptic point of the lower and upper
resonance respectively. The maximum of $\tilde{\omega}(\tilde{H})$
in between $\tilde{H}_{sl}$ and $\tilde{H}_{su}$ is asymptotically
described by the following formula:

\begin{equation}
\tilde{\omega}_m\approx\frac{\pi}{\ln\left(
\tilde{H}_{var}/|\tilde{H}_{sl}-\tilde{H}_{su}|\right)}.
\end{equation}

If we additionally perturb the system in such a way that an
additional time-periodic term of the frequency
$\tilde{\omega}_f\approx \tilde{\omega}_m$ arises in the resonance
Hamiltonian, then the chaotic layers associated with the resonance
separatrices may be connected by chaotic transport even for a
rather small amplitude of the additional perturbation, due to a
scenario similar to the one described in this paper.

There may be various types of such additional perturbation
\cite{webs}. For example, one may {\it add} to $H$ (2) one more
dipole time-periodic perturbation of {\it mixed} frequency (i.e.
$\approx \omega_m+\tilde{\omega}_m$). Alternatively, one may
directly perturb the {\it angle} of the original perturbation by a
{\it low-frequency} perturbation, i.e. the time-periodic term in $H$
(2) is replaced by the term

\begin{eqnarray}
&& -hq\cos(\omega_ft+A\cos(\tilde{\omega}_ft)),
\\
&& \omega_f\approx \omega_m,\quad\quad \tilde{\omega}_f \approx
\tilde{\omega}_m. \nonumber
\end{eqnarray}

Recent physical problems where a similar situation is
relevant are: chaotic mixing and transport in a meandering jet
flow \cite{prants} and reflection of light rays in a corrugated
waveguide \cite{leonel}.

{\bf 4.} If the time-periodic perturbation is {\it multiplicative}
rather than additive,  the resonances become {\it parametric} (cf.\
\cite{Landau:76}). Parametric resonance is more complicated and much
less studied than nonlinear resonance. Nevertheless, the main
mechanism for the onset of global chaos remains the same, namely the
combination of the reconnection between resonances of the same order
and of their overlap in energy with the chaotic layers associated
with the barriers. At the same time, the frequencies of the main
spikes in $h_{gc}(\omega_f)$ may change (though still being related
to $\omega_m$). We consider below the example when the periodically
driven parameter is the parameter $\Phi$ in (1) \cite{37prime_new}.
The Hamiltonian reads as

\begin{eqnarray}
&&
H= p^2/2 +(\Phi-\sin(q))^2/2,
\\
&&
\Phi=\Phi_0+h\cos(\omega_f t), \quad\quad \Phi_0={\rm const}<1.
\nonumber
\end{eqnarray}

\noindent The term $(\Phi-\sin(q))^2/2$ in $H$ (71) may be rewritten
as $(\Phi_0-\sin(q))^2/2
+(\Phi_0-\sin(q))h\cos(\omega_ft)+h^2\cos^2(\omega_ft)/2$. The last
term in the latter expression does not affect the equations of
motion. Thus, we end up with an additive perturbation
$(\Phi_0-\sin(q))h\cos(\omega_ft)$. In the asymptotic limit
$\Phi_0\rightarrow 0$, the $n$th-order Fourier component of the
function $(\Phi_0-\sin(q))$ can be shown to differ from zero only
for the orders $n=2,6,10,...$ Therefore one may expect the main
spikes in $h_{gc}(\omega_f)$ to be at frequencies twice larger
than those for the dipole perturbation (2):

\begin{equation}
\omega_{sp}^{(j)}\approx 2 \omega_{s}^{(j)}\approx
2(2j-1)\omega_m, \quad\quad j=1,2,3,...
\end{equation}

\noindent This well agrees with the results of simulations (Fig. 13).

Moreover, the asymptotic theory for the dipole perturbation may be
immediately generalized to the present case: it is necessary only
to replace the Fourier component of the coordinate $q$ by the Fourier
component of the function $(\Phi_0-\sin(q))$:

\begin{eqnarray}
&& (\Phi_0-\sin(q))_n= \left\{_{0\quad \quad {\rm at} \quad n\neq
2(2j-1),}^{\frac{4}{\pi n}\quad  {\rm at} \quad n= 2(2j-1),} \right.
\\
&& j=1,2,3,..., \nonumber
\\
&& \Phi_0\rightarrow 0 \nonumber
\end{eqnarray}

\noindent (cf. Eq. (12) for $q_n$). We obtain:

\begin{eqnarray}
&& \omega_{sp0}\equiv \omega_{sp0}^{ \left(\frac{n+2}{4}
\right)}=n\frac{\pi}{2\ln \left(\frac{4{\rm e}}{\Phi_0} \right) },
\\
&& h_{sp0}\equiv
h_{sp0}^{\left(\frac{n+2}{4}\right)}=n\frac{c\pi}{8}\frac{\Phi_0}{\ln
\left(\frac{4{\rm e}}{\Phi_0} \right) }, \nonumber
\\
&& n=2,6,10,..., \qquad \Phi_0\rightarrow 0, \nonumber
\end{eqnarray}

\noindent where $c$ is given in Eqs. (50) and (51).

For $\Phi_0=0.2$, Eq. (74) gives, for the 1st spike, values
differing from the simulation data by about $3\%$ for the frequency
and by about $10\%$ for the amplitude. Thus, the lowest-order
formulas accurately describe the 1st spike even for a moderately
small $\Phi$.

\begin{figure}[tb]
\includegraphics[width = 5. cm]{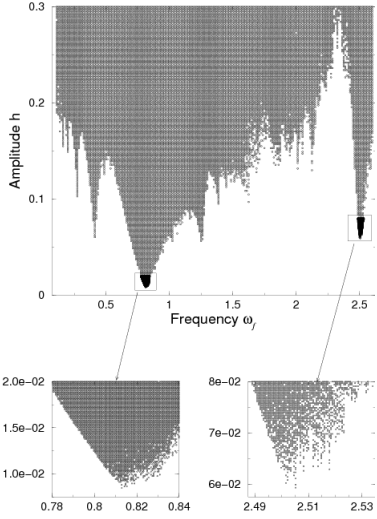}
\caption{The diagram analogous to that one in Fig.4 but for the
system (71) (with $\Phi_0=0.2$). }
\end{figure}

{\bf 5.} One more generalization relates to {\it
multi-dimensional} Hamiltonian systems with two or more saddles
with different energies: the perturbation may not
necessarily be time-periodic, in this case. The detailed analysis will be
done elsewhere.

Finally, we point out some analogy between the facilitation of
global chaos onset described in our work and the so called
stochastic percolation in 2D Hamiltonian systems described in
\cite{percolation} : the merging of internal and external
chaotic zones is also relevant there, like in our case. However,
both the models and the underlying mechanisms are very different.
Namely, the problem studied in \cite{percolation} is not of the
zero-dispersion type; and the two-dimensionality is inherently
important.

Let us turn now to a few detailed examples of applications of the
facilitation of global chaos onset.

\subsection
{Electron gas in a magnetic superlattice, spinning pendulum, cold atoms in an optical lattice}

The first application relates to a classical electron gas in a
magnetic superlattice
\cite{oleg98,oleg99,Oleg12,Oleg10,Shmidt:93,shepelyansky}, where the
electrons may be considered as non-interacting quasi-particles
moving on a plane perpendicular to the magnetic field spatially
periodic in one of the in-plane directions (we denote it as
$x$-direction). Then the electron motion in the x-direction is
described by the Hamiltonian (1) in which $q$ and $p$ are the scaled
electron coordinate $x$ and generalized momentum $p_x$ respectively,
while the parameter $\Phi$ is proportional to the reciprocal
amplitude of the external magnetic field, $B^{-1}$, and to the
generalized momentum $p_y$ in the second (perpendicular to $x$)
in-plane direction: see for details \cite{pr,oleg98,oleg99}. Note
that $p_y$ remains constant during the motion \cite{oleg98,oleg99}.

If an AC electric field is applied in the $x$-direction, then the
model (2)-(1) becomes relevant. The {\it DC-conductivity} in the
$x$-direction is proportional to the fraction of electrons that can
take part to the unbounded motion in the $x$-direction. This
fraction, in turn, significantly grows as the range of energies
involved in the unbounded chaotic transport increases
\cite{oleg98,oleg99}.

If electrons move in vacuum \cite{vacuum}, then it may be possible
to inject a beam of electrons which possess the same velocity. In
this case, the parameter $\Phi$ in the model (1) has some certain
value, so that the results obtained in the previous sections are
directly applicable. The spikes in $h_{gc}(\omega_f)$ mean a drastic
increase of the DC-conductivity occurring at a very weak amplitude
of the AC electric field. The frequency of the AC field should be
close to one of the spike frequencies. The effect is especially
pronounced for the 1st spike i.e. when $\omega_s^{(1)} \approx \pi /
[2\ln(4{\rm e}/\Phi)]$.

If the electron motion takes place in a semiconductor
\cite{oleg98,oleg99,Oleg12,Oleg10}, then the velocity in the
$y$-direction is necessarily {\it statistically distributed}. The
same concerns the parameter $\Phi$ then. This might seem to smear
the effect: cf. \cite{oleg98} where the {\it conventional} scenario
of the onset of global chaos was exploited. However, in the case of
the \lq\lq zero-dispersion'' scenario suggested in our paper, it
typically should not be so. Indeed, a statistical distribution of
the velocity typically decreases {\it exponentially sharply} as the
velocity exceeds some characteristic value $v_c$: for high
temperature $T$, the Boltzmann distribution of energy is relevant
and, therefore, $v_c\propto \sqrt{T}$; for low temperatures, the
Fermi distribution is relevant and, therefore, $v_c\propto
\sqrt{E_F}$ where $E_F$ is the Fermi energy. On the other hand, in
the range of small $\Phi$, the function $\ln(1/\Phi)$ does not
significantly change even if $\Phi$ changes by a few times. Hence,
if $\Phi_c\ll 1$ (where $\Phi_c$ denotes the $\Phi$ value
corresponding to $v_c$), then the frequency $\omega_s^{(1)}$ of the
partial (i.e. for a given value of $v_y$) 1st-order spike is nearly
the same for most of the velocities in the relevant range $v_y
\stackrel {<}{\sim} v_c$, and it is approximately equal to
$\omega_c\equiv \pi /[2\ln(4{\rm e}/\Phi_c)]$. Similarly,
$h_s^{(1)}\sim h_c\approx \omega_c\Phi_c/25$. Thus, as a function of
$\omega_f$  for fixed $h\stackrel{>}{\sim}h_c$, the DC-conductivity
should have a sharp maximum for $\omega_f \approx \omega_c$.

If the parameter $\Phi$ is time-dependent (e.g. if the external
magnetic field has an AC component or/and there is an AC electric
force perpendicular to the $x$-direction), then the applications may
be similar, with the only difference that the values of $\omega_f$
and $h$ in the minima of the main spikes differ from those for the
additive perturbation, see Fig. 13 and the related discussion.

The results of the present paper may also be of direct relevance
for a pendulum spinning about its vertical axis \cite{andronov},
provided the friction is small. The periodic driving may be easily
introduced mechanically, or electrically if the pendulum is
electrically charged, or magnetically if the pendulum includes a
ferromagnetic material.

Finally, we mention in this subsection that potentials similar to
$U(q)$ (1), i.e. periodic potentials with two different-height
barriers per period, may readily be generated for cold atoms by
means of optical lattices \cite{37prime}. The dissipation may be
suppressed by means of the detuning from the atomic resonance
\cite{37prime}. Then the results of our paper are also of direct
relevance to such systems.

\subsection{ Noise-induced escape}

Consider the noise-induced escape over a potential barrier in the
presence of a non-adiabatic periodic driving. For a moderately weak
damping, such a driving decreases the activation barrier due to the
resonant pushing the system in the range of resonant energies
\cite{drsv}. If the damping is even smaller, the decrease of the
activation barrier becomes larger, due, however, to a different
mechanism, typically related to the chaotic layer associated with
the separatrix of the unperturbed system \cite{soskin2000,pr}. The
lower energy boundary of the layer is smaller than the potential
barrier energy, so it is sufficient that the noise pulls the system
right to this boundary (rather than to the very top of the potential
barrier), after which the system may escape over the barrier purely
dynamically. If the eigenfrequency as a function of the energy
possesses a local maximum, then the effect may be even more
pronounced \cite{iric,pr}: the decrease of the activation barrier
may become comparable to the potential barrier at unusually small
amplitudes of the driving, provided the driving frequency is close
to the extremal eigenfrequency. One of the main mechanisms of the
latter effect is closely related to phenomena discussed in the
present paper. In the case of escape over {\it two} barriers of
different heights, the effect should become even more pronounced due
to the mechanism responsible for the spikes of $h_{gc}(\omega_f)$
studied in the present paper. If the potential is periodic, e.g.
like in optical lattices \cite{37prime}, the effect may lead to a
drastic acceleration of the spatial diffusion.

\subsection{Stochastic web}

Our results may be applied to the stochastic web formation
\cite{chernikov1,chernikov2,chernikov3,Zaslavsky:1991}. If a
harmonic oscillator is perturbed by a plane wave whose frequency is
equal to the oscillator eigenfrequency or its multiple, then the
perturbation plays two roles \cite{chernikov2,Zaslavsky:1991}. On
one hand, due to the resonance with the oscillator, it
transforms the structure of the phase space of the oscillator,
leading to an infinite number of cells divided by a unique separatrix.
It has the form of a web of an infinite radius. On the other hand, the
perturbation \lq\lq dresses" this separatrix by an exponentially
narrow chaotic layer (it is sometimes called \lq\lq stochastic''
layer). Such a web-like layer is called {\it stochastic web}. It may
lead to chaotic transport of the system for rather long
distances both in coordinate and in energy.

In case when either the resonance is not exact or/and the
unperturbed oscillator possesses some nonlinearity, the perturbation
generates {\it many} separatrices embedded into each other
\cite{chernikov3,Zaslavsky:1991} rather than one single infinite
web-like separatrix. Then a significant chaotic transport in energy
may arise only if the magnitude of the perturbation exceeds some
critical value corresponding to the overlap of chaotic layers
associated, at least, with two neighbouring separatrices. And,
still, the transport in  energy remains limited since
%the distance between the separatrices grows while
the width of the chaotic layer around each separatrix sharply
decreases as the energy increases. It can be shown \cite{webs} that
some types of additional time-periodic perturbation lead to a
low-frequency dipole perturbation of the resonance system (cf. the
paragraph preceding Eq. (70)). The structure of separatrices in the
reduced system possesses properties similar to that of the system
considered in the present paper. Indeed, in the region between the
separatrices, the resonance system performs regular oscillations,
and the frequency of such oscillations, as function of the value of
the resonance Hamiltonian, is equal to zero at each of the
separatrices. Thus, it necessarily possesses a local maximum between
energies corresponding to any two neighbouring separatrices, like in
the case considered in the present paper. If the additional
perturbation has an optimal frequency related to one of these local
maxima, then the overlap of chaotic layers associated with
neighbouring separatrices is greatly facilitated, similarly to the
case considered in the present paper. Moreover, the local maximum of
the eigenfrequency changes from pair to pair of separatrices {\it
weakly}, so that if the magnitude of the auxiliary perturbation
exceeds the critical value even slightly the simultaneous overlap
between {\it many} chaotic layers may occur. Then, the distance of
the chaotic transport in energy greatly increases.

Similar applications are relevant for the so called homogeneous
(sometimes called periodic) stochastic webs
\cite{chernikov1,Zaslavsky:1991} and many other web-like stochastic
structures \cite{Zaslavsky:1991}.

Beside classical systems, stochastic webs may arise in quantum
systems too. It was recently demonstrated, both theoretically
\cite{Fromhold_PRL} and experimentally \cite{Fromhold_Nature}, that
the stochastic web may play a crucial role in quantum electron
transport in semiconductor superlattices subjected to stationary
electric and magnetic fields. Due to the spatial periodicity with a
period of about a few nanometers, the system possesses narrow
minibands in the electron spectrum. It turns out that the
description of the electron transport in the lowest miniband may be
approximated by the model of a classical harmonic oscillator driven
by a plane wave. The role of the harmonic oscillator is played by
the cyclotron motion while the wave arises due to the interplay
between the cyclotron motion and Bloch oscillations. If the
cyclotron and Bloch frequencies are commensurate, then the phase
space of such a system is threaded by a stochastic web. This gives
rise to the delocalization of electron orbits, that leads in turn to
a strong increase of the conductivity
\cite{Fromhold_PRL,Fromhold_Nature}. However, this effect occurs
only when the ratio between the electric and magnetic fields lies in
the {\it exponentially narrow} regions corresponding to nearly
integer ratios between the Bloch and cyclotron frequencies. The
results of the present work suggest a method for a significant
increase of the width of the relevant regions. If the cyclotron and
Bloch frequencies are not exactly commensurate, then the stochastic
web does not arise: rather a set of embedded separatrices arises
provided the effective wave amplitude is sufficiently large. As
discussed in the previous paragraph, even a rather weak
time-periodic driving \cite {Fromhold_footnote} of the optimal
frequency may significantly increase the area of the phase space
involved in the chaotic transport. This may provide for an effective
control of the electron transport in such a system and may be used
for developing electronic devices that exploit the intrinsic
sensitivity of chaos (cf. \cite{Fromhold_Nature}). A similar effect
may be used also to control transmission through other periodic
structures, e.g. ultra-cold atoms in optical lattices
\cite{Fromhold_Nature_20,Fromhold_Nature_21,Fromhold_Nature_29} and
photonic crystals \cite{Fromhold_Nature_30}.

\section {CONCLUSIONS}

We have developed a new {\it general approach} for the treatment of
separatrix chaos. This has allowed us to create a self-contained
theory for the drastic facilitation of the global chaos onset
between adjacent separatrices of a ID Hamiltonian system subject to
a time-periodic perturbation. Both the new approach and the theory
of the facilitation are closely interwoven in our paper but, at the
same time, each of these two items is relevant even on its own. That
is why we summarize them separately.

{\bf I.} {\it The new approach for separatrix chaos}.

The approach is based on the separatrix map analysis which uses the
characteristic property of the dependence of the frequency of
eigenoscillation on energy in the vicinity of a separatrix: the
frequency keeps nearly a {\it constant} value even if the deviation
of the energy from the separatrix {\it strongly varies} within a
given scale of the deviation. Due to this, the separatrix map
evolves along the major part of the chaotic trajectory in {\it
regular-like} way. The deviation of the chaotic trajectory from the
separatrix may vary along the regular-like parts of the trajectory
in a much wider range than along the irregular parts.

In the case of resonant perturbation, we match the separatrix map
analysis and the resonant Hamiltonian approximation. This allows us
in particular to find the boundaries of the chaotic layers in the
{\it phase space}, which well agrees with computer simulations (Fig.
8). The latter theory has been successfully applied by us to the
problem of the global chaos onset in the double-separatrix case,
which is summarized in the item II below. Other {\it applications}
and {\it generalizations} of the approach include in particular the
following.

\begin {itemize}

\item
%The approach
It may be applied to the conventional {\it single-separatrix} case.
In particular, our theory predicts that the {\it maximal width} of
the separatrix chaotic layer in energy is typically $\gg h$, in
contrast with former theories
\cite{Chirikov:79,lichtenberg_lieberman,Zaslavsky:1991,zaslavsky:1998,zaslavsky:2005}
which assume that the maximal width is $\sim h$.

\item
%The approach
It allows to analyze the {\it transport} within the
%chaotic
layer.

\item
%The approach
It may be generalized for a {\it non-resonant} perturbation and for
a {\it higher} dimension.

\end {itemize}

{\bf II.} {\it The facilitation of the global chaos onset.}

We have considered in details the characteristic example of a
Hamiltonian system possessing {\it two or more separatrices},
subject to a time-periodic perturbation. The frequency $\omega$ of
oscillation of the unperturbed motion necessarily possesses a {\it
local maximum} $\omega_m$ as a function of energy $E$ in the range
between the separatrices. It is smaller than the frequency
$\omega_0$ of eigenoscillation in the stable state of the
Hamiltonian system by a factor

\begin{equation}
R\sim \ln \left(\frac{1}{\Phi}\right), \quad\quad \Phi\equiv
\frac{\Delta U}{U} \, ,
\end{equation}

\noindent where $\Delta U$ is the difference of  the
separatrices energies, while $U$ is the difference between the upper
separatrix energy and the stable state energy.

If $\Phi\ll 1$, the function $\omega(E)$ is close to $\omega_m$ for
most of the energy range between the separatrices: in the asymptotic
limit $\Phi\rightarrow 0$, $\omega(E)$ approaches a {\it
rectangular} form. Besides, the amplitude $q_n$ of the $n$th Fourier
harmonic of the oscillation asymptotically approaches a {\it
non-small} constant value in the whole energy range between
separatrices. These two properties are responsible for most of the
characteristic features of the global chaos onset in between the
separatrices. The most striking one is a {\it drastic facilitation}
of the global chaos onset when the perturbation frequency $\omega_f$
approaches $\omega_m$ or its multiples: the perturbation amplitude
$h_{gc}$ required for global chaos possesses, as a function of the
perturbation frequency $\omega_f$, deep spikes close to $\omega_m$
or its multiples.

On the basis of the theory for the boundaries of the chaotic layers,
we have developed a {\it self-consistent} asymptotic theory for the
{\it spikes} in the vicinity of the minima. In particular, the {\it
explicit} asymptotic expressions for the very minima are given in
Eqs. (52) and (55). The minimal amplitude $h_{gc}$ is smaller than
the typical $h_{gc}$ for $\omega_f$ beyond the close vicinity of
$\omega_m$ by a factor $\sim 10 R\sim 10\ln(1/\Phi)$. The theory
well agrees with the simulation results.

 We have also found the mechanisms responsible for the spike wings (Figs. 9, 11). The
theory well fits the simulations (Fig. 10). The asymptotic shape of
the spike is {\it universal}: it is described by Eq. (67) (Fig. 12).

The facilitation of the global chaos onset may have the following
{\it applications} in particular:

\begin {itemize}

\item drastic increase of the {\it DC-conductivity} of a 2D
electron gas in a 1D magnetic superlattice;

\item significant decrease of the {\it activation barrier} for
noise-induced escape over double/multi-barrier structures, that may
lead to a drastic acceleration of the {\it diffusion} in periodic
structures;

\item strong facilitation of the {\it stochastic web} formation.
\end{itemize}

\section{ACKNOWLEDGMENTS}

The work was partly supported by INTAS Grants 97-574 and 00-00867,
by the Convention between the Institute of Semiconductor Physics and
Pisa University for 2005-2007, and by the ICTP Program for
Short-Term Visits. We are grateful to
%Dominique
D. Escande,
%James
J. Meiss and G. Zaslavsky for discussions. SMS and RM acknowledge
the hospitality of Pisa University and Institute of Semiconductor
Physics respectively.

\appendix
\section{Separatrix map analysis}

The chaotic layers of the system (2) associated with the
separatrices of the unperturbed system (1) are described here by
means of the separatrix map. To derive the map, we follow the method
described in \cite{Zaslavsky:1991}, but the analysis of the map
significantly differs from existing ones
\cite{lichtenberg_lieberman,Zaslavsky:1991,zaslavsky:1998,zaslavsky:2005,treshev}.
It constitutes the
%{\it first ever}
method allowing to calculate chaotic layer boundaries in the {\it
phase space} (rather than just in energy). It also allows to
quantitatively describe the {\it transport} within the layer in a
manner different from existing ones (cf. \cite{treshev,vered} and
references therein).

We present below a detailed consideration of the lower chaotic layer
while the upper layer may be considered similarly and we present
just the results for it.

\begin{figure}[tb]
\includegraphics[width = 6.0 cm]{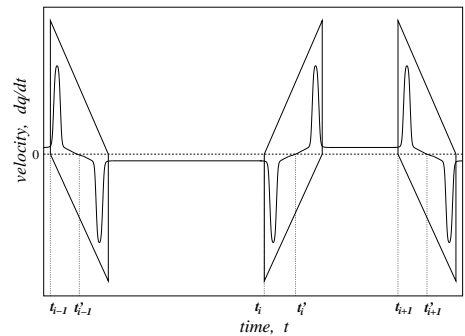}
\caption{Schematic example of the time dependence of the
velocity of the perturbed system (thick solid line) in the case
when the energy of motion varies in the close vicinity of the top of the lower
potential barrier. The dashed line marks
the zero level of the velocity. Pulses of the velocity are schematically
singled out by the
parallelograms (drawn by a thin solid line). The two sequences of time
instants $(...,t_{i-1}, t_i, t_{i+1}, ...)$ and
$(...,t_{i-1}^{\prime}, t_i^{\prime}, t_{i+1}^{\prime}, ...)$
correspond to beginnings and centers of the pulses, respectively.}
\end{figure}

\subsection {Lower chaotic layer}.

\subsubsection {Separatrix map}

The typical function $\dot{q}(t)$ for the trajectory close to the
inner separatrix (the separatrix corresponding to the lower
potential barrier) is shown in Fig. 14. One can resolve pulses in
$\dot{q}(t)$. Each of them consists of two approximately
antisymmetric spikes \cite {pulse}. The pulses are separated by
intervals during which $|\dot{q}|$ is relatively small. Generally
speaking, successive intervals differ of each other. Let us
introduce the pair of variables $ \, E \, $ and $ \, \varphi $:

\begin{equation}
  E \equiv H_0, \quad \varphi\equiv \omega_ft+\varphi_a \, ,
\end{equation}

\noindent where the constant $\varphi_a$ may be chosen
arbitrarily.

The energy $E$ changes only during the pulses of $\dot{q}(t)$ and
remains nearly unchanged during the intervals between the pulses,
when $|\dot{q}(t)|$ is small \cite{Zaslavsky:1991}. We assign
numbers $ \, i \, $ to the pulses and introduce the sequences of
$(E_i,\varphi_i)$ corresponding to initial instants of pulses
$t_i$. In such a way, we obtain the following map (cf.
\cite{Zaslavsky:1991}):

\begin{eqnarray}
&&
E_{i+1}=E_i+\Delta E_i,
\\
&&
\varphi_{i+1}=\varphi_i+\frac{\omega_f\pi(3- {\rm sign}(E_{i+1}-E_b^{(1)}))
}{2\omega(E_{i+1})},
\nonumber
\\
&&
\Delta E_i\equiv h\int_{\hbox{\small{{\it i\,}th pulse}}} \!\!\!\!\!
{\rm d}t\;\dot{q}(t)\cos(\omega_ft),
\nonumber
\end{eqnarray}

\noindent
where $ \, \int_{\hbox{\small{{\it i\,}th pulse}}} \, $
means integration over the $i$th pulse.

Before deriving
a more explicit expression for $\Delta E_i$,
we make two following remarks.

1. Let us denote with $t_i^{\prime}$ the instant within the $i$th
pulse when $\dot{q}$ is equal to zero (Fig. 14). The function
$\dot{q}(t-t_i^{\prime})$ is an odd function within the $i$th pulse
and it is convenient to transform the cosine in the integrand in
$\Delta E_i$ (A2) as

\begin{eqnarray}
&&
\cos(\omega_ft)\equiv\cos(\omega_f(t-t_i^{\prime})+\omega_ft_i^{\prime})\equiv
\nonumber
\\
&&
\cos(\omega_f(t-t_i^{\prime}))
\cos(\omega_ft_i^{\prime})-
\sin(\omega_f(t-t_i^{\prime}))
\sin(\omega_ft_i^{\prime}),
\nonumber
\end{eqnarray}

\noindent and to put $ \varphi_a = \omega_f(t_i^{\prime}-t_i) $, so
that $ \varphi_i\equiv \omega_ft_i^{\prime} $.

2. Each pulse of $\dot{q}$ contains one positive and one negative
spike. The first spike can be either positive or negative. If $E$
changes during the given $n$th pulse so that its value at the end
of the pulse is {\it smaller} than $E_b^{(1)}$, then the first
spikes of the $i$th and $(i+1)$st pulses have the {\it same} signs.
On the contrary, if $E$ at the end of the $i$th pulse is {\it
larger} than $E_b^{(1)}$, then the first spikes of the $i$th and
$(i+1)$st pulses have {\it opposite} signs. Note that Fig. 14
corresponds to the case when the energy remains above $E_b^{(1)}$
during the whole interval shown in the figure. This obviously
affects the sign of $\Delta E_i$, and it may be explicitly accounted
for in the map if we introduce a new discrete variable $\sigma_i=\pm 1$
which characterizes the sign of $\dot{q}$ at the beginning of a
given $i$th pulse,

\begin{equation}
   \sigma_i\equiv {\rm sign} (\dot{q}(t_i)) \ ,
\end{equation}

\noindent
and changes from pulse to pulse as

\begin{equation}
\sigma_{i+1}=\sigma_i \, {\rm sign}(E_b^{(1)}-E_{i+1}) \ .
\end{equation}

\noindent With account taken of the above remarks, we can rewrite
the map (A2) as follows:

\begin{eqnarray}
E_{i+1}&=&E_i+\sigma_ih\epsilon^{(low)}\sin(\varphi_i),
\\
\varphi_{i+1}&=&\varphi_i+\frac{\omega_f\pi(3- {\rm
sign}(E_{i+1}-E_b^{(1)})) }{2\omega(E_{i+1})}, \nonumber
\\
\sigma_{i+1}&=&\sigma_i \, {\rm sign}(E_b^{(1)}-E_{i+1}), \nonumber
\\
\epsilon^{(low)}&\equiv &\epsilon^{(low)}(\omega_f)= \nonumber
\\
&=& -\sigma_i \int_{i{\rm th}\;{\rm pulse}} {\rm
d}t\;\dot{q}(t-t_i^{\prime})\sin(\omega_f(t-t_i^{\prime}))
\approx\nonumber
\\
%\quad\quad\quad\quad\quad\quad\quad\quad
&\approx & -2\sigma_i \int_{t_i^{\prime}}^{t_{i+1}}{\rm
d}t\;\dot{q}(t-t_i^{\prime})\sin(\omega_f(t-t_i^{\prime})).
\nonumber
\end{eqnarray}

The map similar to (A5) was introduced for the first time in
\cite{ZF:1968}, and it is often called as the Zaslavsky separatrix
map. Its mathematically rigorous derivation may be found e.g. in the
recent major mathematical review \cite{treshev}. The latter review
describes also generalizations of the Zaslavsky map as well as other
types of separatrix maps. The analysis presented below relates
immediately to the Zaslavsky map but it is hoped to be possible to
generalize it for other types of the separatrix maps too.

The variable $\epsilon^{(low)}$ introduced in (A5) will be
convenient for the further calculations since it does not depend on
$i$ in the lowest-order approximation. A quantity like
$\delta_l\equiv h |\epsilon^{(low)}|$ is sometimes called the {\it
separatrix split} \cite{zaslavsky:1998} since it is conventionally
assumed that the maximal deviation of energy on the chaotic
trajectory from the separatrix energy is of the order of $\delta_l$
\cite{lichtenberg_lieberman,Zaslavsky:1991,zaslavsky:1998,zaslavsky:2005}.
Though we shall also use this term, we emphasize that the maximal
deviation may be much larger.

Dynamical chaos appears in the separatrix map (A5)
%due to the divergence of $|{\rm d}\omega/{\rm d}E|$ as $E$ approaches
%$E_b^{(1)}$.
because $\omega(E\rightarrow E_b^{(1)})\rightarrow 0$. Various
heuristic criteria were suggested for the estimate of the chaotic
layer width in energy
\cite{lichtenberg_lieberman,Zaslavsky:1991,zaslavsky:1998,zaslavsky:2005}.
Frequencies relevant to our problem are much smaller than the
reciprocal width of the spikes of $\dot{q}(t)$. For such
frequencies, all these criteria
\cite{lichtenberg_lieberman,Zaslavsky:1991,zaslavsky:1998,zaslavsky:2005}
give:

\begin{equation}
|E-E_b^{(1)}|\sim\frac{\omega_f}{\omega_0}h\left|\epsilon^{(low)}\right|,\quad
\omega_f\ll\omega_0\approx 1,
\end{equation}

\noindent where $\omega_0$ is the frequency of eigenoscillation at
the bottom of the potential well.

The estimate (A6) was used in our earlier theory \cite{prl2003}. But
we found later that, for the case of small $\omega_f$, the
aforementioned criteria
%\cite{lichtenberg_lieberman,Zaslavsky:1991,zaslavsky:1998}
were {\it insufficient}, so that the estimate (A6) was incorrect
\cite{to_be_published} (cf. also \cite{E&E:1991,prl2005}). Moreover,
to search a uniform width of the layer is incorrect in cases like
ours, where the width strongly depends on the angle. At the same
time, the {\it lowest-order} formulas for the spike minimum
$(h_s,\omega_s)$ are not affected by this, so that the results of
\cite{prl2003} (with only the lowest-order formulas) are correct.
Still, the higher-order corrections (quite significant for
$h_s^{(j)}$ if $\Phi$ is moderately small) would be incorrect if
they were calculated on the basis of the estimate (A6). Besides, the
paper \cite{prl2003} did not address the intriguing question: why
does even a small excess of $h$ over $h_{gc}(\omega_f)$ result in
the onset of chaos in a large part of the phase space between the
separatrices, despite the fact that the width of the chaotic layers
associated with the nonlinear resonances is {\it exponentially
small} for $h=h_{gc}(\omega_f)$? The analysis of the separatrix map
presented below resolves these important problems.

In the adiabatic limit $\omega_f\rightarrow 0$, the excess of the
upper boundary $E_{cl}^{(1)}$ of the lower layer over the lower
barrier $E_b^{(1)}$ does not depend on angle and equals $2\pi h$
\cite{to_be_published} (cf. also \cite{E&E:1991}). But $\omega_f$
relevant for the spike of $h_{gc}(\omega_f)$ cannot be considered
as an adiabatic frequency, despite its smallness, because it is
close to $\omega_m$ or to its multiple while all energies at the
boundary lie in the range where the eigenfrequency is also close
to $\omega_m$:

\begin{eqnarray}
&& \omega_f\approx(2j-1)\omega_m\approx
(2j-1)\omega(E_{cl}^{(1)}),
\\
&& j=1,2,3,... \nonumber
\end{eqnarray}

The validity of (A7) (confirmed by the results) is {\it crucial} for
the description of the layer boundary in the relevant case.
%Remarkably, this description is quantitative, unlike the
%conventional description of the non-adiabatic case (cf.
%\cite{lichtenberg_lieberman,Zaslavsky:1991,zaslavsky:1998,zaslavsky:2005}).

\subsubsection {Separatrix split}

Let us explicitly evaluate $\epsilon^{(low)}$. Given
that the energy is close to $E_b^{(1)}$, the velocity
$\dot{q}(t-t_i^{\prime})$ in $\epsilon^{(low)}$ (A5) may be
replaced by the corresponding velocity along the separatrix
associated with the lower barrier,
$\dot{q}_s^{(low)}(t-t_i^{\prime})$, while the upper limit in the
integral may be replaced by infinity. Besides, in the asymptotic limit
$\Phi\rightarrow 0$, the interval between spikes within the pulse
becomes infinitely long \cite{pulse} and, therefore, only short
($\sim\omega_0^{-1}$) intervals corresponding to the spikes
contribute to the integral in $\epsilon^{(low)}$ (A5). In the
scale $\omega_f^{-1}$, they may be considered just as two \lq\lq
instants'':

\begin{equation}
t_{sp}^{(1,2)}-t_i^{\prime}\approx\pm\frac{\pi}{4\omega_m}
,\quad
\Phi\rightarrow
0.
\end{equation}

\noindent In the definition of $\epsilon^{(low)}$ (A5), we
substitute the argument of the sine by the corresponding constants
for the positive and negative spikes respectively:

\begin{eqnarray}
&&
\epsilon^{(low)}\approx 2\sin
\left(\frac{\pi\omega_f}{4\omega_m}
\right)
\int_{\rm positive \, spike}{\rm
d}t\;\dot{q}_s^{(low)}(t-t_i^{\prime})
\nonumber
\\
&&
\quad\quad
%\quad
\approx 2\pi\sin
\left(\frac{\pi\omega_f}{4\omega_m}
\right),
\\
&&
\Phi\rightarrow
0.
\nonumber
\end{eqnarray}

\noindent In the derivation of the first equality in (A9), we have
also taken into account that the function
$\dot{q}_s^{(low)}(x)$ is odd. In the derivation of the second equality in (A9), we
have taken into account that the right turning point of the
relevant separatrix is the top of the lower barrier and the
distance between this point and the left turning point of the
separatrix approaches $\pi$ in the limit $\Phi\rightarrow 0$.

For the frequencies relevant to the minima of the spikes of $h_{gc}(\omega_f)$,
i.e. for $\omega_f=\omega_s^{(j)}\approx(2j-1)\omega_m$, we
obtain:

\begin{eqnarray}
&& \epsilon^{(low)}(\omega_s^{(j)}) \approx 2 \pi \sin
\left((2j-1)\frac{\pi}{4} \right) =\sqrt{2}\pi (-1)^{ \left[
\frac{ 2j-1}{ 4 } \right] },\nonumber
\\
&& j =  1,2,3,...,\quad \Phi\rightarrow 0.
\end{eqnarray}

For moderately small $\Phi$, it is better to use the more accurate
formula:

\begin{equation}
\epsilon^{(low)}(\omega_f)= 2\int_0^{\infty}{\rm
d}t\;\dot{q}_s^{(low)}(t)\sin(\omega_ft),
\end{equation}

\noindent where the instant $t=0$ corresponds to the turning point
of the separatrix to the left from the lower barrier, i.e.
$\dot{q}_s^{(low)}(t=0)=0$ while $\dot{q}_s^{(low)}>0$ for all
$t>0$. The dependence $\left|\epsilon^{(low)}(\omega_f)\right|$ by
Eq. (A11) is shown for $\Phi=0.2$ in Fig. 15(a). For small
frequencies, the asymptotic formula (A9) well fits the formula
(A11).

\begin{figure}[tb]
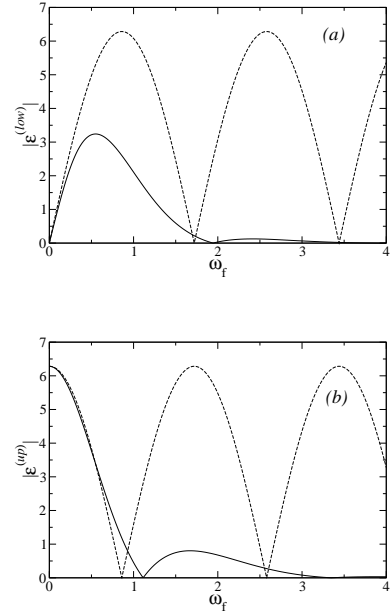

\includegraphics[width = 5. cm]{prefig15a.eps}
\vskip 0.75 cm
\includegraphics[width = 5. cm]{prefig15b.eps}
\vskip 0.5 cm \caption{ The theoretical estimates for the normalized
separatrix split (for $\Phi=0.2$) as a function of the perturbation
frequency, for the lower and upper layers in (a) and (b)
respectively. The solid lines are calculated by Eqs. (A11) and (A43)
(for (a) and (b) respectively) while dashed lines represent
asymptotic expressions (A9) and (A44) respectively. }
\end{figure}

\subsubsection {Dynamics of the map}

Consider the {\it dynamics} of the map (A5), when
$\omega_f$ is close to the spikes minima: $\omega_f\approx
n\omega_m$ where $n\equiv 2j-1$ while $j=1,2,3,\ldots$. Let the
energy at the step $i=-1$ be
%infinitesimally close
equal to $E_b^{(1)}$. The trajectory passing through the state with
this energy is chaotic
%\cite{Chirikov:79,lichtenberg_lieberman,Zaslavsky:1991}
since $(\omega(E))^{-1}$ diverges as $E\rightarrow E_b^{(1)}$ and,
therefore, the angle $\varphi_{-1}$ is not correlated with the
angle on the previous step $\varphi_{-2}$. The quantity
$\sigma_{-1}$ is not correlated with $\sigma_{-2}$ either. Thus,
$\sin(\varphi_{-1})$ may take any value in the range
$[-1,1]$ and $\sigma_{-1}$ may equally take the values 1 or -1.
Therefore, the energy may change on the next step by an arbitrary
value in the interval
$[-h|\epsilon^{(low)}|,h|\epsilon^{(low)}|]$. Thus,
$E_0-E_b^{(1)}$ may have a positive value $\sim
h|\epsilon^{(low)}|$ \cite{except}. Then, the approximate equality
$n\omega(E_0)\approx \omega_m$ holds, provided the value of $h$ is
from the relevant range. Allowing for this and recalling that we are
interested only in those realizations of the map such that
$E_0>E_b^{(1)}$, the relevant realization of the map
$i=-1\;\rightarrow\; i=0$ may be written as:

\begin{eqnarray}
E_{0} & =&E_b^{(1)}+\sigma_{-1}h\epsilon^{(low)}\sin(\varphi_{-1})=
\nonumber
\\
& =&E_b^{(1)}+h|\epsilon^{(low)}\sin(\varphi_{-1})|, \nonumber
\\
\varphi_{0} & \approx &\varphi_{-1}+n\pi,\nonumber
\\
\sigma_0 &&=-\sigma_{-1}.
\end{eqnarray}

One may expect that the further evolution of the map will
approximately follow, for some time, the trajectory of the system
(4) with the initial energy $E_0$ (A12), and with an arbitrary
$\varphi_{-1}$ and initial slow angle $\tilde{\psi}$ somehow related
to $\varphi_0\approx \varphi_{-1}+n\pi$. Let us prove this
explicitly.

Consider two subsequent iterations of the map (A5): $2i\rightarrow
2i+1$ and $2i+1\rightarrow 2i+2$ with an arbitrary $i\ge 0$. While
doing this, we shall assume the validity of (A7) (it will be
clarified below when this is true) from which it follows that: (i)
$\omega(E_{k+1})\approx \omega(E_k)$, (ii)
$\varphi_{k+1}-\varphi_{k}\approx n\pi\equiv (2j-1)\pi$. It will
follow from the results that the neglected corrections are small in
comparison with the characteristic scales of the variation of $E$
and $\varphi$ (cf. the conventional treatment of the nonlinear
resonance dynamics
\cite{Chirikov:79,lichtenberg_lieberman,Zaslavsky:1991,zaslavsky:1998,zaslavsky:2005,pr}).
Besides, it follows from (A5) that, while the energy remains above
the barrier energy,
%the discrete variable
$\sigma_{k}$ oscillates,
so that $\sigma_{2i}=\sigma_{0}$ and $\sigma_{2i+1}=-\sigma_{0}$.
Then,

\begin{eqnarray}
E_{2i+1} & =&E_{2i}+\sigma_0h\epsilon^{(low)}\sin(\varphi_{2i}),
\nonumber
\\
\varphi_{2i+1} &
=&\varphi_{2i}+\frac{\omega_f}{\omega(E_{2i+1})}\pi\approx \nonumber
\\
& \approx & \varphi_{2i}+n\pi
+\pi\frac{\omega_f-n\omega(E_{2i})}{\omega(E_{2i})},
\end{eqnarray}

\begin{eqnarray}
E_{2i+2}&=&E_{2i+1}-\sigma_0h\epsilon^{(low)}\sin(\varphi_{2i+1})=\nonumber
\\
&
=&E_{2i+1}+\sigma_0h\epsilon^{(low)}\sin(\varphi_{2i+1}-n\pi)\approx\nonumber
\\
&\approx & E_{2i}+\sigma_02h\epsilon^{(low)}\sin(\varphi_{2i}),
\nonumber
\\
\varphi_{2i+2}&=&\varphi_{2i+1}+\frac{\omega_f}{\omega(E_{2i+2})}\pi\approx\nonumber
\\
&\approx &\varphi_{2i}+2\pi n
+2\pi\frac{\omega_f-n\omega(E_{2i})}{\omega(E_{2i})}
\end{eqnarray}

\noindent (the second equality in the map for $E_{2i+2}$ takes into
account that $n$ is odd so that
$\sin(\varphi-n\pi)=-\sin(\varphi)$).

The quantity $\varphi_{2i+2}-\varphi_{2i}-2\pi n$ is small, so
the map $2i\rightarrow 2i+2$ (A14) may be approximated by
differential equations for $E_{2i}$ and
$\tilde{\varphi}_{2i}\equiv \varphi_{2i}- 2\pi ni $:

\begin{eqnarray}
&& \frac{{\rm d}E_{2i}}{{\rm
d}(2i)}=\sigma_0h\epsilon^{(low)}\sin(\tilde{\varphi}_{2i}),
\nonumber
\\
&& \frac{{\rm d}\tilde{\varphi}_{2i}}{{\rm d}(2i)}=
\frac{\pi}{\omega(E_{2i})}(\omega_f-n\omega(E_{2i})),
\\
&& \tilde{\varphi}_{2i}\equiv \varphi_{2i}- 2\pi ni. \nonumber
\end{eqnarray}

Let us (i) use for $\epsilon^{(low)}$ the asymptotic formula (A10),
(ii) take into account that the increase of $i$ by 1 corresponds to the
increase of time by
$\pi/\omega(E)$, and (iii) transform from the variables
$(E,\tilde{\varphi})$ to the variables $(I,\tilde{\psi}\equiv
n\pi(1-\sigma_0)/2-\tilde{\varphi})$. Equations (A15) reduce then
to:

\begin{eqnarray}
&& \frac{{\rm d}I}{{\rm
d}t}=-h\sqrt{2}(-1)^{\left[\frac{n}{4}\right]}\sin(\tilde{\psi}),
\nonumber
\\
&& \frac{{\rm d}\tilde{\psi}}{{\rm d}t}= n\omega-\omega_f,
\\
&& \tilde{\psi}\equiv n\pi\frac{1-\sigma_0}{2}-\tilde{\varphi},
\quad\quad n\equiv 2j-1. \nonumber
\end{eqnarray}

Equations (A16) are identical to the equations of motion of the
system (4) in the lowest-order approximation, i.e. to the equations
(14) where $q_n$ is replaced by its asymptotic value (12) and the
last term in the right-hand part of the second equation is
neglected, being of higher order in comparison with the term
$n\omega-\omega_f$.

Apart from the formal identity of Eqs. (A16) and (14),
$\tilde{\psi}$ in (A16) and $\tilde{\psi}$ in (14) are identical
to each other. Necessarily $t_i^{\prime}$
corresponds to a turning point (see Fig. 14) while the corresponding
$\psi$ is equal to $2\pi i$ or $\pi+2\pi i$ for the right and left
turning point respectively (see (4)) i.e. $\psi=2\pi
i+\pi(1-\sigma_i)/2$, so that $\tilde{\psi}_{(14)}\equiv
n\psi-\omega_ft=n\pi(1-\sigma)/2-\tilde{\varphi}\equiv
\tilde{\psi}_{(A16)}$.

The relevant initial conditions for (A16) follow from (A12) and from
the relation between $\tilde{\psi}$ and $\varphi$:

\begin{equation}
I(0)= I(E=E_b^{(1)}+h\sqrt{2}\pi|\sin(\tilde{\psi}(0))|),
\end{equation}

\noindent while $\tilde{\psi}(0)\equiv
n\pi(1-\sigma_0)/2-\varphi_{0}$ may be an arbitrary angle from the
ranges where

\begin{equation}
(-1)^{[n/4]}\sin(\tilde{\psi}(0))<0.
\end{equation}

For moderately small $\Phi$, it is better to use the  more accurate
dynamic equations (14) instead of (A16) and the more accurate
initial value of action instead of (A17):

\begin{equation}
I(0)=I(E=E_b^{(1)}+\delta_l|\sin(\tilde{\psi}(0))|), \quad\quad \delta_l\equiv
h|\epsilon^{(low)}|,
\end{equation}

\noindent with $\epsilon^{(low)}$ calculated by (A11).

We name the quantity $\delta_l|\sin(\tilde{\psi})|$ the {\it
generalized separatrix split} (GSS) for the lower layer. Unlike the
conventional separatrix split $\delta_l$ \cite{zaslavsky:1998}, it
is {\it angle-dependent}. The curve
$I(\tilde{\psi})=I(E=E_b^{(1)}+\delta_l|\sin(\tilde{\psi})|)$ may be
called then the GSS curve for the lower barrier and denoted as
$I_{\rm GSS}^{(l)}(\tilde{\psi})$.

Finally, let us investigate an important issue: whether the
transformation from the discrete separatrix map (i.e. (A13) and
(A14)) to the differential equations (A15) is valid for the very
first step and, if it is so, for how long it is valid after that.
The transformation is valid as long as $\omega(E_k)\approx
n\omega_f$ i.e. as long as $E_k$ is not too close to the barrier
energy $E_b^{(1)}$. At the step $k=0$, the system stays at the GSS
curve, with a given (random) angle $\tilde{\psi}(0)$ from the range
(A18). Thus, at this stage, the relation (A7) is certainly valid
(for the relevant range of $h$ and for an angle a little away from
the narrow vicinity of the multiples of $\pi$).

The change of energy at the next step is positive too:

\begin{eqnarray}
E_1-E_0 &&\equiv
\sigma_0h\epsilon^{(low)}\sin(\tilde{\varphi}_{0})\approx\nonumber
\\
&&\approx
-\sigma_{-1}h\epsilon^{(low)}\sin(\tilde{\varphi}_{-1}-n\pi)=
\nonumber\\
&& = \sigma_{-1}h\epsilon^{(low)}\sin(\tilde{\varphi}_{-1})\equiv
E_0-E_{-1}>0.\nonumber
\end{eqnarray}

\noindent This may  also be interpreted as a consequence of the
first equation in (A16) and of the inequality (A18).

Hence, (A7) is valid at the step $k=1$ too. Similarly, one can show
that $E_2-E_1>0$, etc. Thus, the transformation
(A13,A14)$\rightarrow$(A15) is valid at this initial stage indeed,
and the evolution of $(E,\tilde{\varphi})$ does reduce to the
resonant trajectory (14) with an initial angle from the range (A18)
and the initial action (A19). This lasts until the resonant
trajectory meets the GSS curve in the adjacent $\pi$ range of
$\tilde{\psi}$ i.e. at $t$ such that the state
$(I(t),\tilde{\psi}(t))$ satisfies the conditions:

\begin{equation}
I(t)=I_{\rm GSS}^{(l)}(\tilde{\psi}(t)),\nonumber \quad\quad
[\tilde{\psi}(t)/\pi]-[\tilde{\psi}(0)/\pi]=1.
\end{equation}

\noindent At this instant, the absolute value of the change of
energy $E_k$ in the separatrix map (A13) is equal to $E_k-E_b^{(1)}$
(just because the state belongs to the GSS curve) but the sign of
this change is negative since the sign of $\sin(\varphi_k)$ is
opposite to the sign of $\sin(\varphi_0)$. Therefore, at the step
$k+1$, the system gets to the very separatrix, and the regular-like
evolution stops: at the next step of the map, the system may either
again get to the GSS curve with a new (random) angle from the range
(A18) and start a new regular-like evolution as described above; or
it may get to the similar GSS curve {\it below} the barrier and
start an analogous regular-like evolution in the energy range below
the barrier, until it stops in the same manner as described above,
etc.

This approach makes it possible to describe all
features of the transport within the chaotic layer. Their detailed
description will be done elsewhere while, in the present context,
it is most important to describe the {\it upper outer boundary} of
the layer.

\subsubsection {Boundary of the layer}

We may now analyze the evolution of the boundary of the layer as $h$
grows. Some of the evolution stages are illustrated by Figs. 8, 9
and 16.

\begin{figure}[tb]
\includegraphics[width = 6.5 cm]{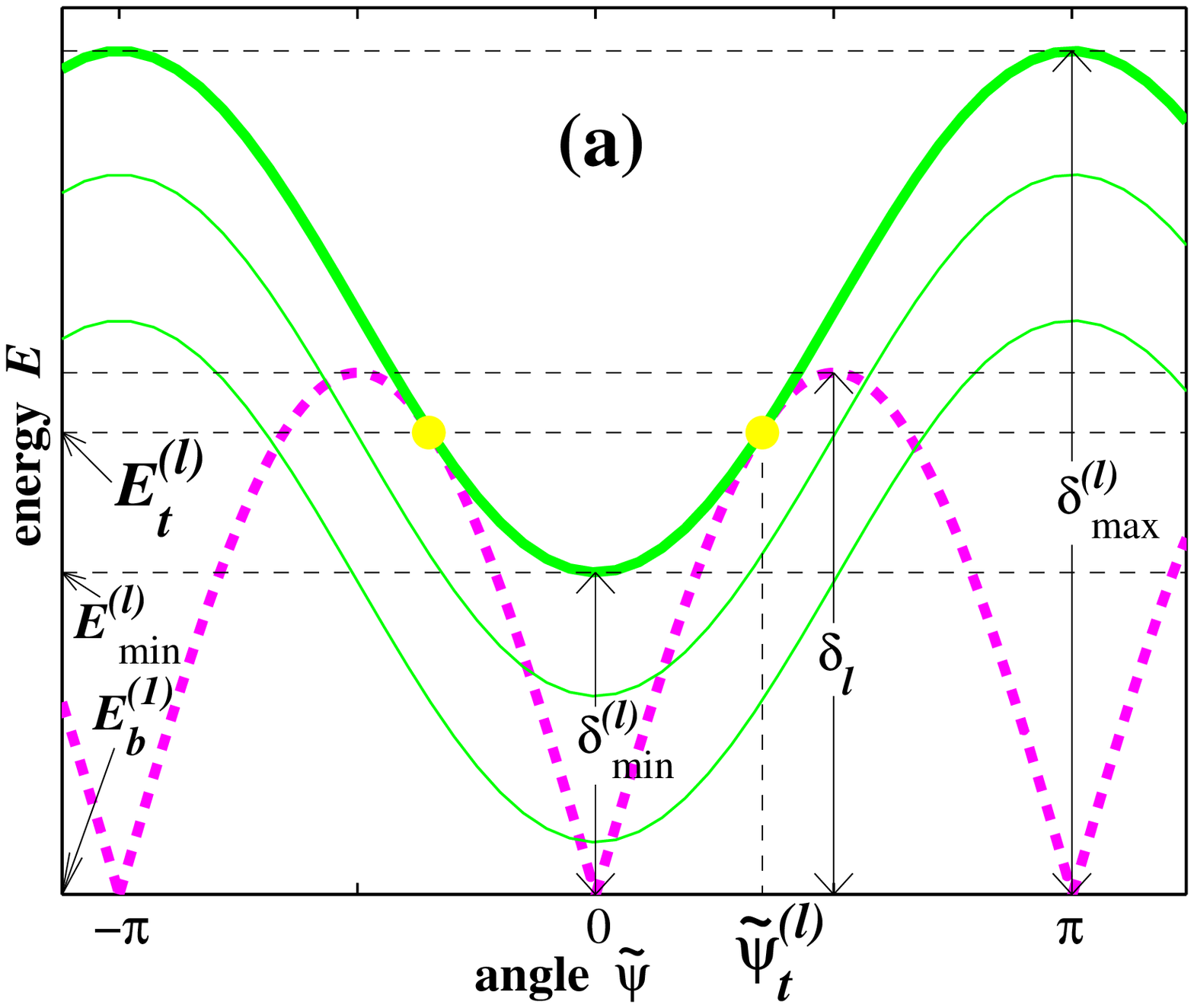}
\includegraphics[width = 6.5 cm]{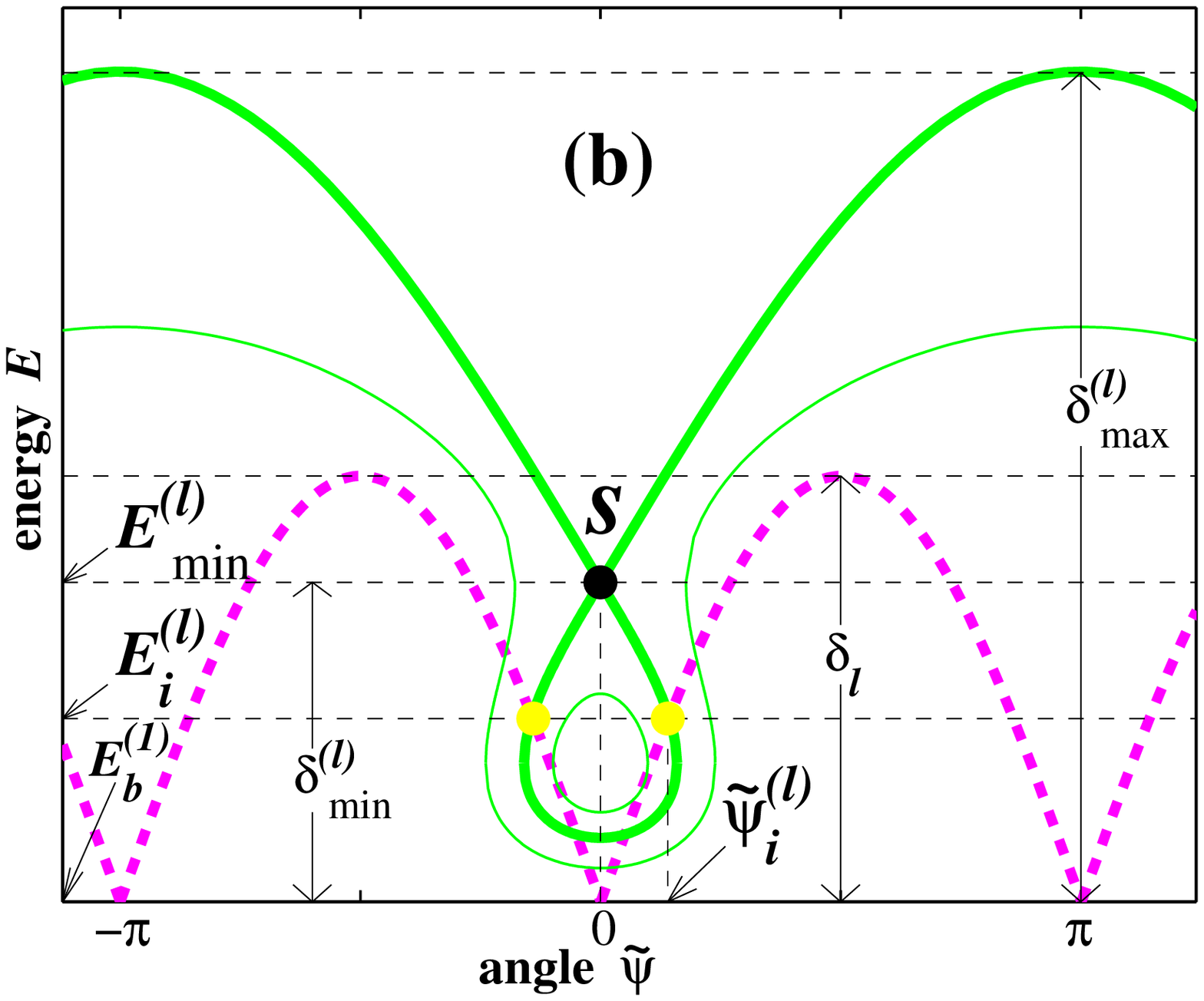}
\caption{A schematic figure illustrating the formation of the
boundary of the lower chaotic layer for $h<h_{cr}^{(l)}(\omega_f)$
in the ranges of $\omega_f$ relevant to (a) odd, and (b) even
spikes. The dashed magenta line shows the GSS curve in the
energy-angle plane: $E(\tilde {\psi})=E_{\rm GSS}^{(l)}(\tilde
{\psi})\equiv E_b^{(1)}+\delta_l|\sin(\tilde {\psi})|$. Green lines
show examples of those trajectories (14) which have points in common
with the GSS curve. One of such trajectories (14) (shown by the {\it
thick} green line) forms the upper boundary of the lower chaotic
layer: in (a), it is the trajectory  {\it tangent} to the GSS curve;
in (b), it is the upper part of the separatrix generated by the
saddle \lq\lq $s$''. Yellow dots indicate the relevant common points
of the GSS curve and the thick green line. They have angles
$\pm\tilde {\psi}_t^{(l)}$ and energy $E_t^{(l)}$ in the case (a),
and angles $\pm\tilde {\psi}_i^{(l)}$ and energy $E_i^{(l)}$ in the
case (b). The minimum and maximum deviation of energy on the
boundary from the barrier energy are denoted as
$\delta_{\min}^{(l)}$ and $\delta_{\max}^{(l)}$ respectively. The
maximum deviation on the GSS curve is equal to $\delta_l$. }
\end{figure}

It follows from the analysis carried out in the previous subsection
that {\it any} state (in the $I-\tilde {\psi}$ plane) lying beyond
the GSS curve but belonging to any trajectory following the
equations (14) which possesses common points with the GSS curve
belongs to the chaotic layer: the system starting from such a state
will get, sooner or later, to the separatrix, where the chaotization
will necessarily occur. Therefore,  the  {\it upper boundary} of the
chaotic layer coincides with the trajectory following the equations
(14) with the initial action (A19) and an initial angle $\tilde
{\psi}(0)$ from the range (A18) such that the trajectory deviates
from the barrier energy more than a trajectory (14)-(A18)-(A19) with
any other initial angle does. There may be only two topologically
different options for such a trajectory: either it is the trajectory
{\it tangent} to the GSS curve, or it is the separatrix trajectory
which {\it intersects} the GSS curve (some schematic examples are
shown in Figs. 16(a) and 16(b) respectively; some real calculations
are shown in Figs. 8 and 9).

\vskip 0.2cm

\hskip 2.5cm {\it 1. Relatively small $h$}

\vskip 0.2cm

Consider first values of $h$ which are large enough for the
condition (A7) to be satisfied (the explicit criterion will be given
in (A31)) but which are smaller than the value $h_{cr}^{(l)}\equiv
h_{cr}^{(l)}(\omega_f)$ determined by Eq. (41) (its meaning is
explained below). The further analysis within this range of $h$
differs for the ranges of $\omega_f$ relevant to {\it odd} and {\it
even} spikes, and we consider them separately.

\vskip 0.2cm

\hskip 3.cm {\it A. Odd spikes}

\vskip 0.2cm

The relevant frequencies are:

\begin{eqnarray}
&& \omega_f\approx n\omega_m, \quad\quad n\equiv 2j-1, \nonumber\\
&& j=1,3,5, \ldots
\end{eqnarray}

Let us seek the state $\{I_t^{(l)},\tilde{\psi}_t^{(l)}\}$ (with
$\tilde{\psi}_t^{(l)}$ within the range $]0,\pi[$) where the
resonant trajectory curve is {\it tangent} to the GSS curve. With
this aim, we equal both the actions and the derivatives of both
curves. The equality of actions immediately yields $I_t^{(l)}$ via
$\tilde{\psi}_t^{(l)}$: $I_t^{(l)}\equiv I(E=E_t^{(l)})=I_{\rm
GSS}^{(l)}(\tilde{\psi}_t^{(l)})$. The derivative along the GSS
curve is obtained by  differentiation of $I_{\rm
GSS}^{(l)}(\tilde{\psi})$.
%(note that ${\rm d}I/{\rm d}E=\omega^{-1}$).
The derivative
%${\rm d}I/{\rm d}\tilde{\psi}$
along a resonant trajectory can be
found dividing the first dynamic equation in
(14) by the second one. Substituting the expression of $I_t^{(l)}$
via $\tilde{\psi}_t^{(l)}$ into the equality of the derivatives, we obtain
a closed equation for $\tilde{\psi}_t^{(l)}$, and its solution
immediately
gives us the relevant $\tilde{\psi}(0)$:

\begin{eqnarray}
&&\left[
|\epsilon^{(low)}|\cos(\tilde{\psi}_{t}^{(l)})
\left(
1-\frac{\omega_f}{n\omega(E)}-h\frac{{\rm d}q_n(E)}{{\rm d}E}
\cos(\tilde{\psi}_{t}^{(l)})
\right)+
\right.
\nonumber
\\
&&\left. +q_n(E)\sin(\tilde{\psi}_{t}^{(l)})
\right]_{E=E_t^{(l)}}=0,
\\
&& E_t^{(l)}\equiv
E_b^{(1)}+h|\epsilon^{(low)}|\sin(\tilde{\psi}_{t}^{(l)}),
\nonumber\\
&& \tilde{\psi}_{t}^{(l)}\in [0,\pi],
 \quad\quad
n\equiv 2j-1, \quad\quad j=1,3,5,\ldots, \nonumber
\\
&& \tilde{\psi}(0)=\tilde{\psi}_{t}^{(l)}. \nonumber
\end{eqnarray}

A careful analysis of the phase space structure shows that, in the
present case (i.e. when $h<h_{cr}^{(l)}(\omega_f)$ while $j$ is
odd), there is no separatrix of the resonant Hamiltonian (4) which
would both intersect the GSS curve and possess points above the
tangent trajectory \cite{irrelevant}. Thus, for this range of $h$,
the outer boundary of the chaotic layer is formed by the trajectory
following the dynamical equations (14) with the initial angle by
(A22) and initial action (A19) (Fig. 16(a)).

Let us find the lowest-order solution of Eq. (A22). We neglect the
term $1-\omega_f/(n\omega(E))$ (the result will confirm the validity
of this) and use the lowest-order expression for the relevant
quantities: namely, Eqs. (A10) and (12) for $\epsilon^{(low)}$ and
$q_n$ respectively, and the lowest-order expression for ${\rm
d}q_n/{\rm d}E$ which can be derived from Eq. (11):

\begin{eqnarray}
&& \frac{{\rm d}q_n(E)}{{\rm d}E}=(-1)^{\left[\frac{n}{4}\right]}
\frac{\pi}{4\sqrt{2}\left(E-E_b^{(1)}\right)\ln\left(\Phi^{-1}\right)},
\nonumber
\\
&& n\equiv 2j-1,\quad\quad E-E_b^{(1)}\ll \Phi\rightarrow 0.
\end{eqnarray}

Then Eq. (A22) reduces to the following equation

\begin{equation}
\tan^2(\tilde{\psi}_{t}^{(l)})=
\frac{n\pi}{8\ln\left(\Phi^{-1}\right)}.
\end{equation}

The lowest-order solution of (A24) in the range $]0,\pi[$ reads as

\begin{equation}
\tilde{\psi}_{t}^{(l)}=(-1)^{\left[\frac{n}{4}\right]}\sqrt{\frac{n\pi}{8\ln(1/\Phi)}}
+\pi\frac{1-(-1)^{\left[\frac{n}{4}\right]}}{2}.
\end{equation}

\noindent It follows from the definition $E_t^{(l)}$ (A22) and from
(A25) that the lowest-order expression for $E_t^{(l)}-E_b^{(1)}$
reads as

\begin{equation}
E_t^{(l)}-E_b^{(1)} = \delta_l\sin(\tilde{\psi}_t^{(l)})  =
\frac{\pi^{3/2}} {2}\frac{h}{\sqrt{\ln\left(1/\Phi\right)/n}}.
\end{equation}

The next step is to find the {\it minimal} value of the energy on the
boundary of the layer, $E_{\min}^{(l)}$. It follows from the
analysis of the dynamical equations (14) that the corresponding angle
$\tilde{\psi}_{\min}$ is equal to $0$ if ${\rm sign} (q_{2j-1})>0$
(i.e. $j=1,5,9,\ldots$) or to $\pi$ if ${\rm sign} (q_{2j-1})<0$
(i.e. $j=3,7,11,\ldots$): cf. Fig. 8(a). Given that the
Hamiltonian (4) is constant along any trajectory (14) while the
boundary coincides with one of such trajectories, the values of
the Hamiltonian (4) in the states
$\{I(E_{\min}^{(l)}),\tilde{\psi}=\tilde{\psi}_{\min}\}$ and
$\{I_{t}^{(l)},\tilde{\psi}_{t}^{(l)}\}$ should be equal to each
other. In the explicit form, this equality may be written as

\begin{eqnarray}
&&\int_{E_{\min}^{(l)}}^{E_t^{(l)}}{\rm d}E\left(1-\frac{\omega_f}{n\omega(E)}\right)
-
\\
&&
-h\left(q_n(E_t^{(l)})\cos(\tilde{\psi}_t^{(l)})-(-1)^{\left[\frac{n}{4}\right]}q_n(E_{\min}^{(l)})\right)=0.
\nonumber
\end{eqnarray}

Let us find the lowest-order solution of Eq. (A27). Assume that
$E_{\min}^{(l)}$ still belongs to the range of $E$ where
$\omega(E)\approx\omega_m$ (the result will confirm this
assumption). Then the integrand in (A27) goes to zero in the
asymptotic limit $\Phi\rightarrow 0$ and, hence, the integral may be
neglected (the result will confirm the validity of this). The
remaining terms in Eq. (A27) should be treated very carefully. In
particular, it is insufficient to use the lowest-order value (12)
for $q_n$ since it is the difference between $q_n(E_t^{(l)})$ and
$q_n(E_{\min}^{(l)})$ that matters. Moreover, the approximate
equality $q_n(E_t^{(l)})-q_n(E_{\min}^{(l)}) \approx {\rm
d}q_n(E_t^{(l)})/{\rm d}E_t^{(l)}(E_t^{(l)}-E_{\min}^{(l)})$ does
not apply here either since, as it follows from Eq. (A23), the
derivative ${\rm d}q_n(E)/{\rm d}E$ may strongly vary in the range
$[E_{\min}^{(l)},E_{t}^{(l)}]$ if
$(E_t^{(l)}-E_{\min}^{(l)})/(E_{\min}^{(l)}-E_b^{(1)})\stackrel{\sim}{>}1$
(the result will show that it is the case). That is why it is
necessary to use for $q_n$ the more accurate expression (11).
Allowing for the asymptotic expression (A25) of
$\tilde{\psi}_t^{(l)}$ and keeping only the lowest-order terms, one
can finally reduce Eq. (A27) to the relation

\begin{equation}
\ln\left(\frac{E_t^{(l)}-E_b^{(1)}}{E_{\min}^{(l)}-E_b^{(1)}}\right)=
\frac{1}{2}.
\end{equation}

\noindent Substituting here the asymptotic value of $E_t^{(l)}$
(A26), we obtain the final lowest-order expression for the minimal
(along the boundary) deviation of the energy from the barrier:

\begin{eqnarray}
\delta_{\min}^{(l)}&\equiv&
E_{\min}^{(l)}-E_b^{(1)}=(E_t^{(l)}-E_b^{(1)})/\sqrt{{\rm e}}=
\nonumber
\\
&=&\frac{\pi^{3/2}} {2{\rm
e}^{1/2}}\frac{h}{\sqrt{\ln\left(1/\Phi\right)/n}}.
\end{eqnarray}

It is necessary and sufficient that the condition $\omega(E)\approx
\omega_m$ is satisfied at the {\it minimal} and {\it maximal}
energies of the boundary to ensure that the second equality in (A7)
holds true, i.e. that $\omega(E)$ is close to $\omega_m$ for {\it
all} points of the boundary.

At the minimal energy, this condition reads as

\begin{equation}
\omega_m-\omega(E_b^{(1)}+\delta_{\min}^{(l)})\ll\omega_m.
\end{equation}

\noindent Eq. (A30) determines the lower limit of the relevant range
of $h$. The asymptotic form of (A30) is:

\begin{equation}
\frac{\ln\left( \frac{\Phi\sqrt{\ln(1/\Phi)}}{h}\right)}{\ln\left(
1/\Phi \right)}\ll 1.
\end{equation}

\noindent We emphasize that any $h$ of the order of $h_{s0}$ (52)
satisfies this condition. In the asymptotic limit $\Phi\rightarrow
0$, the left-hand part of Eq. (A31) goes to zero.

As for the {\it maximal} energy, it may take values up to the energy
of the lower saddle \lq\lq $sl$'', i.e. $E_{sl}$ (18). Obviously,
(A7) is valid at this saddle, too.

\vskip 0.2cm

\hskip 3.cm {\it B. Even spikes}

\vskip 0.2cm

The relevant frequencies are:

\begin{eqnarray}
&& \omega_f\approx n\omega_m, \quad\quad n\equiv 2j-1, \nonumber\\
&& j=2,4,6, \ldots
\end{eqnarray}

In this case, $q_n(E)$ and ${\rm d}q_n(E)/{\rm d}E$ have different
signs for all $E$ within the relevant range (i.e. where
$\omega(E)\approx \omega_m$, $q_n(E)\approx q_n(E_m)$): cf. (12) and
(A23). Then, in the asymptotic limit $\Phi\rightarrow 0$, Eq. (A22)
for the tangency does not have any solution for
$\tilde{\psi}_t^{(l)}$ in the relevant range \cite{58_prime}. There
may be only solutions very close to some of $\pi$ integers, and the
corresponding energies $E_t^{(l)}$ are very close to $E_b^{(1)}$
i.e. $\omega(E_t^{(l)})\ll \omega_m$: therefore they are irrelevant.

At the same time, unlike in the case of odd spikes, there exists a
saddle with an angle

\begin{equation}
\tilde{\psi}_s^{(l)}=\pi\frac{1-(-1)^{\left[\frac{n}{4}\right]}}{2},
\end{equation}

\noindent while the energy (which may be found as the appropriate
solution of Eq. (15)) lies in the relevant vicinity of the lower
barrier (Fig. 16(b)). In the lowest-order approximation, this saddle
energy reads:

\begin{equation}
E_s^{(l)}\equiv E_b^{(1)}+\delta_s,\quad\quad
\delta_s=\frac{\pi}{2\sqrt{2}}\frac{h}{\ln(\ln(4{\rm e}/\Phi))} \;
.
\end{equation}

This saddle (denoted in Fig. 16(b) as \lq\lq$s$'') generates a
separatrix. Its upper whiskers go to the similar adjacent saddles
(shifted in $\tilde{\psi}$ by $2\pi$). In the asymptotic limit
$\Phi\rightarrow 0$, the upper whiskers are much steeper than the
GSS curve and hence they do not intersect it \cite{58_prime_prime}.
As concerns the lower whiskers, they do intersect the GSS curve and,
moreover, two intersections lie in the relevant energy range (Fig.
16(b)). Let us show this explicitly. Let us write the expression for
the Hamiltonian (4) in the relevant vicinity of the barrier energy
(i.e. where $\omega_m-\omega(E)\ll\omega_m)$, keeping, in the
expression, both the lowest-order terms and the terms of next order
(in particular, we use Eq. (11) for $q_n(E)$ and take into account
that $0<\sqrt{2}-nq_n(E)\ll \sqrt{2}$ for the relevant range of
$E$):

\begin{eqnarray}
&&\tilde{H}(I=I(E=E_b^{(1)}+\delta),\tilde{\psi})=
\nonumber\\
&&=-\frac{n\delta\ln \left(\frac{2\Phi}{\delta}\right)}{2\ln
\left(\frac{4{\rm
e}}{\Phi}\right)}+\left(\omega_f-\frac{n\pi}{2\ln
\left(\frac{4{\rm e}}{\Phi}\right)}\right)\frac{2\Phi}{\pi}\ln
\left(\frac{4{\rm
e}}{\Phi}\right)- \nonumber\\
&&-(-1)^{\left[\frac{n}{4}\right]}h\sqrt{2}\left(1+\frac{n\pi\ln
\left(\frac{2\Phi}{\delta}\right)}{8\ln \left(\frac{4{\rm
e}}{\Phi}\right)}\right)\cos(\tilde{\psi}),\\
&& \omega_m-\omega(E+\delta)\ll\omega_m. \nonumber
\end{eqnarray}

The Hamiltonian $\tilde{H}$ should possess equal values at the
saddle \lq\lq $s$'' and at the intersections of the separatrix and
the GSS curve. Let us denote the angle of the intersection in the
range $]0,\pi[$ as $\tilde{\psi}_i^{(l)}$, and let us denote the
deviation of its energy $E_i^{(l)}$ from $E_b^{(1)}$ as
$\delta_i^{(l)}\equiv\delta_l\sin(\tilde{\psi}_i^{(l)})$.

Assuming that $|\tilde{\psi}_i^{(l)}-\tilde{\psi}_s^{(l)}|\ll 1$
(the result will confirm this) so that
$\cos(\tilde{\psi}_i^{(l)})\approx
(-1)^{[n/4]}(1-(\tilde{\psi}_i^{(l)}-\tilde{\psi}_s^{(l)})^2/2)\approx
(-1)^{[n/4]}(1-(\delta_i^{(l)}/\delta_l)^2/2)\approx
(-1)^{[n/4]}(1-(\delta_i^{(l)}/h)^2/4)$, the equality of the
values of $\tilde{H}$ reads as:

\begin{eqnarray}
&&\frac{n}{2\ln \left(\frac{4{\rm e}}{\Phi}\right)} \left(
\delta_s\ln \left(\frac{2\Phi}{\delta_s}\right)-\delta_i^{(l)}\ln
\left(\frac{2\Phi}{\delta_i^{(l)}}\right)\right)= \nonumber
\\
&& = h\sqrt{2}\frac{n\pi}{8}\frac{\ln
\left(\frac{\delta_s}{\delta_i^{(l)}}\right)}{\ln
\left(\frac{4{\rm
e}}{\Phi}\right)}-\frac{(\delta_i^{(l)})^2}{2\sqrt{2}h}.
\end{eqnarray}

Let us assume that, in the asymptotic limit $\Phi\rightarrow 0$,
$\delta_i^{(l)}\ll \delta_s$ (the result will confirm this). Then
the left-hand part is asymptotically smaller than the first term in
the right-hand part. So, Eq. (A36) implies, in the asymptotic limit,
that the right-hand side equals zero. Expressing $h$ via $\delta_s$
from Eq. (A34), we finally obtain a closed transcendental equation
for $\delta_s/\delta_i^{(l)}$:

\begin{eqnarray}
&&\left(\frac{\delta_s}{\delta_i^{(l)}}\right)^2\ln\left(\frac{\delta_s}{\delta_i^{(l)}}\right)=
\nonumber
\\
&& =\frac{\pi\ln \left(\frac{4{\rm
e}}{\Phi}\right)}{n\left(\ln\left(\ln \left(\frac{4{\rm
e}}{\Phi}\right)\right)\right)^2}\equiv A.
\end{eqnarray}

In the asymptotic limit  $\Phi\rightarrow 0$, the quantity $A$
diverges and, hence, the lowest-order asymptotic solution of Eq.
(A37) reads as

\begin{equation}
\frac{\delta_s}{\delta_i^{(l)}}=\sqrt{\frac{2A}{\ln(A)}}.
\end{equation}

\noindent Substituting here the expression (A34) for $\delta_s$ and
the expression (A37) for $A$, we  obtain:

\begin{equation}
\delta_i^{(l)}=h\frac{1}{4}\sqrt {\frac{n\pi\ln\left(\ln
\left(\frac{4{\rm e}}{\Phi}\right)\right)}{\ln \left(\frac{4{\rm
e}}{\Phi}\right)}}.
\end{equation}

Thus, we have proved the following asymptotic properties of the
separatrix generated by the saddle \lq\lq$s$'': 1) the lower
whiskers of the separatrix do intersect the GSS curve in the
relevant range of $E$ (i.e. where the resonant approximation is
valid), 2) the upper whiskers of the separatrix do {\it not}
intersect the GSS curve (there is no solution of Eq. (A36) in the
range $\delta_i^{(l)}>\delta_s$). The former property confirms the
self-consistence of the asymptotic theory for even spikes; the
latter property means that the {\it upper outer boundary} of the
lower chaotic layer is formed by the {\it upper whiskers of the
separatrix generated by the saddle \lq\lq$s$''}.

Finally, we explicitly note that the minimal (along the boundary)
deviation of energy from the barrier energy occurs occurs right at
the saddle \lq\lq$s$'', i.e.

\begin{equation}
\delta_{\min}^{(l)}=\delta_s.
\end{equation}

\vskip 0.2cm

\hskip 3.cm {\it 2. Relatively large $h$.}

\vskip 0.2cm

%Thus, (A7) is valid for all points of the relevant boundaries of both layers.

As $h$ grows, the boundary of the
%lower
layer raises up while the lower part of the resonance separatrix, on
the contrary, goes down. They reconnect at the critical value of
$h$, $h_{cr}^{(l)}\equiv h_{cr}^{(l)}(\omega_f)$, determined by Eq.
(41), that may be considered as the absorption of the resonance by
the chaotic layer. If $h$ grows further, then the GSS curve and the
resonance separatrix  intersect. As a result, the trajectory
starting from the state with the angle (A22) and action (A19), for
odd spikes, or from the saddle \lq\lq $s$\rq\rq, for even spikes, is
{\it encompassed} by the resonance separatrix. So, it does not form
the outer boundary of the layer anymore. Rather it forms the inner
boundary i.e. the boundary of the main island of the stability
inside the layer, repeated periodically in $\tilde{\psi}$ with a
period $2\pi$ (cf. analogous islands in the upper layer in Fig. 8).
Unless the lower chaotic layer reconnects with the upper one, the
{\it outer} boundary of the lower layer is formed by the upper
%(i.e. above the saddle {\it\lq\lq sl\rq\rq})
part of the {\it resonance separatrix}. The
relevant initial angle $\tilde{\psi}(0)$ on the GSS curve corresponds
to the intersection of the GSS
curve with the resonance separatrix (cf. the analogous situation for the upper layer in Fig. 8).

\subsection {Upper chaotic layer}

The upper chaotic layer may be treated analogously
\cite{footnote_divergence} to the lower layer. We present here only
the results.

Similarly to the lower-layer case, one may consider the ranges of
relatively small $h$ (namely, smaller than $h_{cr}^{(u)}\equiv
h_{cr}^{(u)}(\omega_f)$ determined by Eq. (42)) and relatively large
$h$ (i.e. $h>h_{cr}^{(u)}$). In the former range, the formation of
the boundary occurs in a manner which is, in a sense, opposite to
that for the lower-layer case. For even spikes, the lower outer
boundary is formed by {\it tangency} while, for  odd spikes,
it is formed by the lower part of the {\it separatrix} generated by
the saddle \lq\lq$\tilde{s}$\rq\rq, analogous to the saddle
\lq\lq$s$\rq\rq in the lower-layer case \cite{59_prime}.

So, for even spikes, the angle of tangency
$\tilde{\psi}_t^{(u)}$ is determined by the following equation

\begin{eqnarray}
&&\left[ |\epsilon^{(up)}|\cos(\tilde{\psi}_{t}^{(u)}) \left(
1-\frac{\omega_f}{n\omega(E)}-h\frac{{\rm d}q_n(E)}{{\rm d}E}
\cos(\tilde{\psi}_{t}^{(u)}) \right)- \right. \nonumber
\\
&& \left. -q_n(E)\sin(\tilde{\psi}_{t}^{(u)})
\right]_{E=E_t^{(u)}}=0,
\\
&& E_t^{(u)}\equiv
E_b^{(2)}-h|\epsilon^{(up)}|\sin(\tilde{\psi}_{t}^{(u)})\nonumber
\\
&& \tilde{\psi}_{t}^{(u)}\in\left[0,\pi \right], \quad\quad
n\equiv 2j-1, \quad\quad j=2,4,6,\ldots, \nonumber
\\
&& \tilde{\psi}(0)=\tilde{\psi}_{t}^{(u)},\nonumber
\end{eqnarray}

\noindent
and $\tilde{\psi}_t^{(u)}$ determines the tangency energy:

\begin{equation}
E_t^{(u)}= E_b^{(2)}-h|\epsilon^{(up)}|\sin(\tilde{\psi}_{t}^{(u)}),
\end{equation}

\noindent
where the quantity $\epsilon^{(up)}$ is described by the formula

\begin{equation}
\epsilon^{(up)}(\omega_f)=2\int_0^{\infty}{\rm
d}t\;\dot{q}_s^{(up)}(t)\cos(\omega_ft) \, ,
\end{equation}
where $\dot{q}_s^{(up)}(t)$ is the time dependence of the velocity
along the separatrix associated with the upper barrier and the
instant $t=0$ is chosen so that $q_s^{(up)}(t=0)$ is equal to the
coordinate of the lower barrier while $\dot{q}_s^{(up)}>0$ for $ t
\in [ 0, \infty [$. The dependence
$\left|\epsilon^{(up)}(\omega_f)\right|$ in Eq. (A43) is shown for
$\Phi=0.2$ in Fig. 15(b).

The asymptotic form of Eq. (A43) reads as

\begin{equation}
\epsilon^{(up)}\equiv
\epsilon^{(up)}(\omega_f)=2\pi\cos\left(\frac{\pi\omega_f}{4\omega_m}
\right) \, .
\end{equation}

\noindent For $\omega_f=\omega_s^{(j)}\approx
(2j-1)\omega_m$, Eq. (A43) reduces to

\begin{eqnarray}
&& \epsilon^{(up)}(\omega_s^{(j)}) \approx 2 \pi \cos
\left((2j-1)\frac{\pi}{4} \right) =\sqrt{2}\pi (-1)^{ \left[
\frac{ 2j+1}{ 4 } \right] },\nonumber
\\
&& j =  1,2,3,...,\quad \Phi\rightarrow 0.
\end{eqnarray}

The lowest-order solution of (A41) is given in Eq. (37), so that
$E_{t}^{(u)}$ is approximated by Eq. (38). The maximal energy on the
lower boundary of the layer corresponds to $\tilde{\psi}(t)=\pi$ if
$ j=2,6,10,\ldots $ or  $0$ if $ j=4,8,12,\ldots$ and is determined
by Eq. (39). The asymptotic value of the minimal deviation from the
upper barrier of the energy at the boundary, $\delta_{\min}^{(u)}$,
is given in Eq. (40).

For odd spikes, the boundary is formed by the lower part of the
separatrix generated by the saddle \lq\lq $\tilde{s}$ \rq\rq. The angle of the
saddle is given in Eq. (33) while the deviation of its energy from the barrier
is approximated, to the lowest-order approximation, by Eq. (34).

As $h$ grows, the boundary of the layer goes down while the upper
part of the upper resonance separatrix goes up. They
reconnect at $h=h_{cr}^{(u)}\equiv
h_{cr}^{(u)}(\omega_f)$, as
determined by Eq. (42), that may be considered as the absorption of the
resonance by the chaotic layer.

For larger $h$, the boundary of the layer is formed by the
lower
%(i.e. below the saddle {\it\lq\lq su\rq\rq})
part of the
upper resonance separatrix (Fig. 8), unless the latter intersects the lower
GSS curve (in the latter case, $h_{cr}^{(u)}$ marks the global chaos onset).


\begin{thebibliography}{99}
\bibitem{more} More complicated cases are considered e.g. in
\cite{pr,Howard:95}.
\bibitem{20_prime}
This value of $h$ is so close to the critical value $h_{gc}$ (the
latter is about 0.0049, as seen from Fig. 4) that the blue and
green layers nearly touch each other. That is why we did not draw
a regular trajectory separating the layers.
\bibitem{footnote2}
We still use different colours for the trajectories starting from
$\{I^{(0)}\}$ and $\{O^{(0)}\}$ in order to demonstrate that their
mixing occurs very slowly, which indicates that $h$ is only slightly
above the critical $h_{gc}$.
\bibitem{even} Even harmonics are absent in the eigenoscillation
due to the symmetry of the potential.
\bibitem{barriers} Note that $E_{b}^{(1,2)}\approx 1/2\mp \Phi$ to first order
in $\Phi$.
\bibitem{arbitrary_constant} We omit here and in Eq. (10)
corrections $\sim(\ln(1/\Phi))^{-1}$
since they vanish in the asymptotic limit $\Phi\rightarrow 0$.
\bibitem {crosses} In the Poincar\'{e} sections shown in
Fig. 5, the points which correspond to such stable points of
 equations (14)
are indicated by the crosses.
\bibitem{27} Eq. (20) is the {\it sufficient} (rather than only
necessary) condition for the separatrix reconnection since there
is no any other separatrix which would lie in between the
separatrices generated by the saddles \lq\lq {\it sl}'' and \lq\lq
{\it su}''.
\bibitem{28_prime} With account for the next-order corrections,
the spike minimum $(h_s,\omega_s)$ coincides with
$(h_s^{(l)},\omega_s^{(l)})$ in case of an odd spike or with
$(h_s^{(u)},\omega_s^{(u)})$ in case of an even spike. This occurs
because, in case of odd spikes, $|q_n(E)|$ increases/decreases as
$E$ approaches the relevant vicinity of the upper/lower barrier,
while it is vice versa in case of even spikes. And the larger
$|q_n|$ the farer the resonance separatrix extends: in other words,
the reconnection of the barrier chaotic layer with the resonance
separatrix requires a smaller value of $h$ at the barrier where
$|q_n|$, in the relevant vicinity of the barrier, is larger.
\bibitem{26_prime} For a given $\omega_f\approx\omega_s^{(j)}$, the numerical
  error
on $h_{gc}$ is rather large since the corresponding simulation
involves a very long integration time. We estimate the
relative numerical error of
$h_{gc}(\omega_f\approx\omega_s^{(1,2)})$ to be
$\stackrel{~}{<}5\%$. Note however that this error practically
does not affect the measured values of the frequencies
$\omega_s^{(j)}$ in the minima of the spikes: the error in the
measurement of $h_{gc}$ is systematic while the spikes are very
sharp. We estimate the error of $\omega_s^{(1,2)}$ to be
$\stackrel{~}{<}0.5\%$.
\bibitem{32_prime} For $n>1$, it is also necessary to check if
the solution lies above the line (66). If it does not , then (20)
should be replaced by (66) in the search for the minimum.
\bibitem{29_prime}
Still, this procedure does not pretend to be exaustive since it
takes into account only corrections within the description by means
of the resonant Hamiltonian (4), while ignoring a correction to the
resonant Hamiltonian (4) itself. A derivation of the latter
correction requires the use of the second-order averaging method
\cite{bogmit}, that goes beyond the scope of the present paper. On
the other hand, even our incomplete procedure gives a good agreement
with the simulations. It means that the corrections coming from the
second-order averaging are probably insignificant in comparison with
the corrections of other origins.
\bibitem{general_case} In the general case of an arbitrary potential
$U(q)$, when the explicit expression for $q(E,\psi)$ and
$\omega(E)$ cannot be obtained, these functions can be
calculated numerically.
\bibitem{fluctuations} Provided
$h_{gc}(\omega_f)$ is smoothed over small fluctuations.
\bibitem{37prime_new} In the case of a
2D electron gas in a magnetic superlattice, this may correspond e.g.\
to the time-periodic electric force applied perpendicular to the
direction of the periodic magnetic field \cite{oleg98,oleg99}.
\bibitem{Fromhold_footnote} It may be induced e.g. by means of a
small modulation of the electric field (i.e. an addition of a small
AC component to the DC field).
\bibitem{pulse} Spikes correspond to the motion
over any of the minima of the potential, first in one and then
(after the reflection from one of the upper barriers) in the
opposite direction. If $\Phi$ is small, the spikes within the pulse
are separated by a long interval since the reflection point is
situated close to the top of the upper barrier, where the motion is
slow.
\bibitem{except}
The latter is valid for any $\varphi_{-1}$ except in the close vicinity
of multiples of $\pi$ while the state $E_0,\varphi_0$ (A5) in the
latter range of $\varphi_{-1}$ turns out irrelevant to the
boundary, as shown further down.
\bibitem{irrelevant} For odd numbers $j\geq 3$, there are separatrices
 which lie in
the range of $E$ where $\omega(E)\ll \omega_m$ i.e. much closer to
the barrier than the tangent trajectory: these separatrices are
therefore irrelevant.
\bibitem{58_prime_prime} In case of a {\it moderately} small $\Phi$,
they may intersect the GSS curve. And then, the tangent trajectory
lying above the separatrix necessarily exists, so that the
boundary of the layer is formed by this tangent trajectory.
\bibitem{footnote_divergence}
For any AC-driven spatially periodic Hamiltonian system, the {\it
upper} energy boundary of the layer associated with the unbounded
separatrix diverges in the adiabatic limit $\omega_f\rightarrow 0$
\cite{prl2005}. However, this divergence is not relevant for the
present problem for the following reasons. The lower chaotic layer
relates to the {\it bounded} separatrix while, for the upper
(unbounded) layer, it is the {\it lower} boundary of the layer which
is relevant for the onset of global chaos in between the
separatrices. Moreover, even for the upper boundary of the upper
layer, the divergence is not yet manifested for the driving
parameters $(h,\omega_f)$ in the vicinity of the spikes minima (cf.
\cite{prl2005}).
\bibitem{59_prime} For a {\it moderately} small $\Phi$, the
tangency may exist. The boundary is formed then by the tangent
trajectory rather than by the separatrix: see an example in Fig.
9(c).
\end{thebibliography}
\end{document}